\documentclass[twoside,a4paper,12pt]{report} 
\usepackage{fancyhdr}
\usepackage{graphics}
\newcommand{\mathfrak}{\cal}
\newcommand{\tr}{\mathop{\rm tr}\nolimits}
\newcommand{\Tr}{\mathop{\rm Tr}\nolimits}
\newcommand{\SU}{\mathop{\rm SU}}
\newcommand{\U}{\mathop{\rm {}U}}
\newcommand{\rmd}{{\rm d}}

\newcommand\fverb{\setbox\pippobox=\hbox\bgroup\verb}
\newcommand\fverbdo{\egroup\medskip\noindent%
			\fbox{\unhbox\pippobox}\ }
\newcommand\fverbit{\egroup\item[\fbox{\unhbox\pippobox}]}
\newbox\pippobox

\newcommand{\sla}[1]{{\ooalign{\hfil/\hfil\crcr$#1$}}}
\newcommand{\f}[2]{\frac{#1}{#2}}

\newcommand{\lb}[0]{\left[}  
\newcommand{\rb}[0]{\right]}
\newcommand{\lc}[0]{\left\{} 
\newcommand{\rc}[0]{\right\}}

\def\hatmu{{\hat\mu}}

\def\ga{\gamma}
\def\g5{\gamma_5}

\newcommand{\plb}[3]{Phys. Lett. {\bf B#1} (#2) #3} 
\newcommand{\prl}[3]{Phys. Rev. Lett. {\bf #1} (#2) #3}
\newcommand{\prd}[3]{Phys. Rev. {\bf D#1} (#2) #3}
\newcommand{\npb}[3]{Nucl. Phys. {\bf B#1} (#2) #3}
\newcommand{\npps}[3]{Nucl. Phys. {\bf B}(Proc. Suppl.) 
{\bf #1} (#2) #3}
\newcommand{\ptp}[3]{Prog. Theor. Phys. {\bf #1} (#2) #3}

\newcommand{\jhep}[3]{J. High Energy Phys.{\bf #1} (#2) #3}
\newcommand{\arnps}[3]{Ann. Rev. Nucl. Part. Sci. {\bf #1} 
(#2) #3}
\newcommand{\cmp}[3]{Comm. Math. Phys. {\bf #1} (#2) #3}
\newcommand{\rmp}[3]{Rev. Mod. Phys. {\bf #1} (#2) #3}
\newcommand{\hepth}[1]{{\bf hep-th/#1}}

\newcommand{\heplat}[1]{{\bf hep-lat/#1}}

\begin{document}
\thispagestyle{empty}
{~~~~~~~}
\vspace{18ex}
\begin{center}
{\LARGE CP symmetry and lattice chiral gauge theories}
\vspace{10ex}\\
{\Large Masato Ishibashi\\
Department of Physics, University of Tokyo
\vspace{30ex}\\
{\large A dissertation presented to the faculty 
of University of Tokyo \\
in candidacy for 
the degree of Doctor of Science}
}
\end{center}




\newpage
\abstract{
The CP symmetry is a fundamental discrete symmetry 
in chiral gauge theory. Therefore 
this symmetry is expected to be kept also on the lattice.
However, it has been pointed out by Hasenfratz 
that the chiral fermion action 
in L\"{u}scher's formulation of lattice chiral gauge theory   
is not invariant under CP transformation. In this thesis, 
we first review the method of constructing  
chiral gauge theory on the lattice. 
Then we generalize the analysis of Hasenfratz and show that 
CP symmetry is not manifestly implemented for the local and
doubler-free Ginsparg-Wilson operator 
under rather general assumptions for chiral projection 
operators. We next calculate the fermion generating functional
and  
precisely identify where the effects of 
this CP breaking appear in this formulation. 
We show
that they appear in: (I)~Overall constant phase of the fermion
generating functional. (II)~Overall dimensionful 
constant 
of the fermion
generating functional. (III)Fermion propagator appearing in 
external
fermion lines and the propagator connected to Yukawa vertices. 
The first
effect appears from the transformation of the path integral 
measure and
it is absorbed into a suitable definition of the constant 
phase factor
for each topological sector; in this sense 
there appears no
``CP anomaly''. The second constant arises from the explicit 
breaking in
the action and it is absorbed by the suitable weights with 
which
topological sectors are summed. The last one 
in the propagator is
inherent to this formulation. This breaking emerges 
as an (almost)
contact term in the propagator when the Higgs field, which 
is treated
perturbatively, has no vacuum expectation value. 
In the presence of the 
vacuum expectation value, however, a completely new situation 
arises and
the breaking becomes intrinsically non-local. 
This non-local CP
breaking is expected to persist for a non-perturbative 
treatment of the
Higgs coupling. The basics of 
lattice gauge theory are briefly summarized in Appendix.A.   
}

\pagestyle{fancy}

\tableofcontents

\chapter{Introduction}
Lattice gauge theory has been the most successful tool 
to calculate 
various physical quantities in gauge theory in 
the strong coupling region since it was proposed by 
Wilson in 1974~\cite{Wilsonlattice}. Lattice 
gauge theory is defined on a discrete Euclidean space-time 
which provides a non-perturbative regularization scheme. 
An ultra-violet cutoff is given by the inverse of the lattice 
spacing. 
Lattice gauge theory in finite volume is also a good starting 
point for the constructive quantum field 
theory, since neither ultra-violet divergences nor infrared 
divergences exist there. An important feature of lattice 
gauge theory is to enable the numerical analysis of 
gauge theory with the computer by using the method similar to 
those used for the statistical mechanics systems.  
Until now, various physical quantities in low energy QCD, 
hadron masses, quark masses, string tension and so on, which 
we can not calculate in perturbation theory, have been 
calculated by the numerical method.~\footnote{The values of 
the physical quantities which are calculated by lattice 
theory have been yearly improved. In this thesis we will treat 
the analytical aspects rather than the numerical aspects of 
lattice theory. For the numerical aspects(and also 
for analytical aspects) the proceedings of annual lattice 
conferences are very useful~\cite{Lattice}.}

While gauge fields can be latticized as link variables 
excellently, the problem, which is so-called the fermion 
doubling problem, appears when one tries to 
put fermions on the lattice~\cite{Wilsonfermion}. 
In particular, it was considered to be difficult that one puts 
a massless fermion on the lattice because of the Nielsen-Ninomiya 
no-go theorem~\cite{NNtheorem}. However, the recent progress in 
the treatment of the fermion, which began with 
the domain-wall fermion~\cite{Kaplan,Shamir},  
has enabled us to put the massless 
fermion on the lattice. Dirac operators which describe a 
single massless fermion have been discovered, for example, 
the overlap Dirac operator~\cite{Neuberger:1998fp} derived from 
the overlap formalism~\cite{Narayanan:1993wx,
Randjbar-Daemi:1995sq} and the fixed point action 
through renormalization group approaches.\footnote{
We do not discuss the fixed point action in this thesis. 
See refs.\cite{Hasenfratz:1998ft} 
for further details.}
While these Dirac operators are derived through quite 
different procedures, they satisfy the Ginsparg-Wilson 
relation~\cite{Ginsparg:1982bj}. Moreover, L\"{u}scher pointed out 
that the Ginsparg-Wilson 
relation is equivalent to a lattice chiral 
symmetry~\cite{luscherchiral}. 
See 
refs.~\cite{Niedermayer:1999bi}--\cite{Kikukawa:2001jk} for 
reviews on this progress.
As for the numerical aspects, 
the numerical analysis of the 
weak matrix 
elements concerning the CP 
breaking parameter $\epsilon^\prime/\epsilon$ 
band $\Delta I=1/2$ rule in the Standard Model has made 
significant progress recently 
in the formulation of QCD with the domain-wall 
fermion~\cite{Noaki:2001un}-\cite{Ishizuka:2002nm}, 
though the definite 
results have not been obtained yet.  

The Standard Model which describes the behaviour of elementary 
particles at present is a chiral gauge 
theory~\cite{Peskin}.
In chiral 
gauge theory 
left-handed fermions and right-handed fermions belong to different 
gauge-group representations. 
A peculiar feature of chiral gauge theory is the appearance of  
the gauge anomaly, which is the breaking of gauge symmetry 
by quantum 
effects.\footnote{As a review on chiral gauge theory 
in continuum theory, 
see, for instance, ref.~\cite{anomalyreview}.} 
Whether gauge anomalies appear or not depends on the gauge group and 
the fermion multiplet. The theory with such anomalies has no 
meaning, since such anomalies break the fundamental principles 
in the theory, for example, unitarity of S matrix. Therefore 
we have to consider a theory where all of the possible anomalies 
cancell. In perturbation theory 
it has been shown that 
there exist no gauge anomalies 
to all orders in 
gauge couplings, if gauge anomalies 
cancel at one-loop level. Further conditions, however, may be 
needed to keep  
gauge symmetry 
at the non-perturbative level. For example, the chiral gauge 
theory with 
an odd number of Weyl fermion multiplets  
in SU(2) fundamental representation has no anomalies in  
perturbation theory. At the non-perturbative level, however, 
the global 
anomaly, which is called the Witten 
anomaly~\cite{Wittenanomaly,Wittenanomalyen}, appears and thus 
this theory 
is inconsistent. 

In chiral gauge theory another problem 
is the regularization problem. 
The appearance of 
gauge anomalies implies that the gauge-invariant regularization 
is possible
only when the multiplets of Weyl fermions are anomaly-free.  
Therefore, 
the gauge-invariant regularization should be directly 
related to the 
fermion representation 
of the gauge group. The standard regularization schemes, 
for example, 
dimensional regularization, Pauli-Villars method and so on, 
are not quite convenient to analyze 
the gauge symmetry in chiral gauge theory. 
If a more fundamental gauge-invariant regularization 
scheme is  
established, it may simplify the calculations of 
radiative corrections 
in electroweak processes. 

The recent progress in lattice gauge theory has led us 
also to 
the exciting possibilities for  
constructing non-perturbative chiral gauge theory with exact 
gauge invariance~\cite{Luscher:1999du}-\cite{kikudwf}. 
The construction of lattice 
chiral gauge theory with exact gauge invariance is called 
L\"{u}scher's formulation, which is adopted in this 
thesis.
\footnote{L\"{u}scher's formulation is different from 
overlap formalism of chiral gauge theory only by 
the phase choice of the fermion measure. On the overlap 
formalism, See refs.\cite{Narayanan:1993wx,
Randjbar-Daemi:1995sq, neuchiral}.} 
In L\"{u}scher's formulation with the Ginsparg-Wilson operators, 
anomaly-free $U(1)$ chiral gauge theory 
in finite volume~\cite{Luscher:1999du}
~\footnote{In the overlap formalism for the fermion 
determinant, noncompact chiral U(1) gauge theories also 
have been constructed on the lattice~\cite{neunon}.} and 
electroweak $SU(2)\times U(1)$ 
chiral gauge theory~\cite{Kikukawa:2001kd} in infinite volume 
have been constructed on the lattice. The global anomaly is also 
analyzed on the lattice~\cite{BarCampos, Bar}.
\footnote{See ref.\cite{neuwitten} for an earlier analysis
of the global anomaly in the overlap formalism.} Further, 
in perturbation 
theory chiral gauge theories with exact gauge invariance are 
constructed 
on the lattice to all orders of the gauge coupling 
for general anomaly-free 
gauge group~\cite{Luscher:2000zd}.   
In addition, a programme 
for the 
construction of chiral gauge theories with general  
anomaly-free gauge groups in a non-perturbative manner 
has been proposed~\cite{Luscher:2000un}. 
Therefore, 
the numerical analysis of the Standard Model, such as, 
the investigation of the behaviour of elctroweak phase 
transition at finite temperature, is coming to be possible. 
This construction of lattice chiral gauge theory  
will also deepen our understanding of chiral gauge theory 
in the strong coupling region. 

It has been however pointed out by Hasenfratz that 
the chiral fermion action 
in this formulation is not invariant under CP 
transformation~\cite{Hasenfratz:2001bz}, 
if one uses the ordinary 
lattice chiral projectors~\cite{Narayanan:1998uu,
Niedermayer:1999bi}. CP symmetry 
is a fundamental discrete symmetry in chiral gauge theory and 
hence is expected to be kept on the lattice. This peculiar feature 
of lattice chiral gauge theory may influence the lattice 
analysis of 
the Standard Model which keeps CP symmetry except a phase factor in 
CKM matrix. Therefore it is very important to 
examine the detailed breaking of CP symmetry 
in this formulation. 

In this thesis the breaking of CP symmetry in lattice 
chiral gauge theory will be examined on the basis of 
two papers~\cite{ffis,fiss}. 
In Chapter 2 we will review  
the formulation of lattice chiral gauge theory in detail,  
following refs.~\cite{Luscher:1999du,Luscher:2000un}.
In Chapter 3, Hasenfratz's analysis 
is extended, utilizing more general projection operators~\cite{ffis}. Then 
we will show that the CP breaking in chiral fermion action 
persists under the standard
CP transformation, even if one utilizes a general class 
of Ginsparg-Wilson operators 
which are local and free of species doublers and utilizes rather 
general chiral 
projections. The CP
breaking is thus 
regarded as an inherent feature of this class of formulation.  
Therefore, we will next examine this CP breaking 
at the quantum level by calculating the fermion 
generating functional~\cite{fiss}. 
Then we will show that
the effects of CP violation 
emerge only at: (I)~Overall constant phase
of the fermion generating functional. (II)~Overall dimensionful 
constant 
of the fermion generating functional. (III)~Fermion propagator 
appearing
in external fermion lines and the propagator connected to 
Yukawa vertices. Our result is summarized in
eq.~(\ref{eq:fourxtwentyone}) for pure chiral gauge theory 
and, when
there is a Yukawa coupling, in eq.~(\ref{eq:sixxeight}). 
The first two
constants above depend only on the topological sectors 
concerned and the
problem is reduced to the choice of weights with which 
various
topological sectors are summed. This problem, which is not 
particular to
this formulation, is thus fixed by a suitable choice of 
weight factors.
The last effect in the fermion propagator, 
on the other hand, can not be removed. 
When the Higgs field has no 
expectation
value,  
it emerges as an (almost) contact term.
However, in the presence of the Higgs expectation 
value, this breaking becomes intrinsically non-local.  
In Chapter 4, we will summarize our result and give a few 
remarks.

\chapter{Lattice chiral gauge theory}\label{latchiralgauge}
In this chapter L\"{u}scher's formulation of 
lattice chiral gauge theory is 
presented~\cite{Luscher:1999du,Luscher:2000un}.
We first discuss the lattice Dirac operator used
in this formulation. Next, Weyl fermions are defined, 
utilizing 
the chiral projection operators. The definition
of the fermion integration measure is 
non-trivial, however, and there exists the freedom of 
the phase of the measure. The conditions for fixing the phase, so 
that the formulation
satisfies gauge invariance and all other fundamental 
principle
, is summarized in L\"{u}scher's reconstruction theorem. 
     
\section{Lattice Dirac operator}\label{latdiracop}
We first consider the Dirac fermion and 
the projection to the left-handed components
is discussed in the next section. 

We consider
a general form of the Ginsparg-Wilson relation.
In terms of the hermitian 
operator $H$:~\footnote{We assume
the $\gamma_5-$hermicity:
\begin{eqnarray}
D^\dagger = \gamma_5 D\gamma_5,
\label{ghermicity}
\end{eqnarray}
where $D$ is the Dirac operator and write the fermion action 
as follows, 
\begin{eqnarray}
&&S_F=a^4\sum_x \bar{\psi}(x)D\psi(x).
\end{eqnarray}
}
\begin{equation}
   H=a\gamma_5D=aD^\dagger\gamma_5=H^\dagger,
\label{oph}
\end{equation}
this relation is written as
\begin{equation}
\label{generalGW}
   \gamma_5H+H\gamma_5=2H^2f(H^2),
\end{equation}
where $f(H^2)$ is a regular function of $H^2$ and
$f(H^2)^\dagger=f(H^2)$. For simplicity, we assume 
that $f(x)$ is
monotonous and non-decreasing for~$x\geq0$.\footnote{This general 
Ginsparg-Wilson relation 
is also written as follows, 
\begin{eqnarray}
&& \gamma_5 D + D\gamma_5=2aDf(a^2D^\dagger D)\gamma_5 D,
\end{eqnarray}
by using $D$.}  
The simplest choice
$f(H^2)=1$ corresponds to the conventional Ginsparg-Wilson
relation~(\ref{GW}).
The properties of $H$ for $f(H^2)=H^{2k}$ with 
a positive integer $k$ have been 
investigated~\cite{Fujikawa:2000my}-\cite{fujipro}.
As an 
important consequence of eq.(\ref{generalGW}), one has
\begin{eqnarray}
 && \gamma_5H^2=(\gamma_5H+H\gamma_5)H-H(\gamma_5H+H\gamma_5)
   +H^2\gamma_5=H^2\gamma_5,
\end{eqnarray}
which implies $DH^{2}=H^{2}D$.    
The combination
\begin{equation}
   \Gamma_5=\gamma_5-Hf(H^2),
\label{eq:twoxfive}
\end{equation}
has a special role due to the property
\begin{equation}
   \Gamma_5H+H\Gamma_5=0.
\label{eq:twoxsix}
\end{equation}
Moreover, the above defining relation~(\ref{generalGW}) is written 
in a variety of ways
such as  
\begin{eqnarray}
\label{varietyGW}
&&\gamma_{5}\Gamma_{5}\gamma_{5}D + D\Gamma_{5}=0,\nonumber\\
&&\gamma_{5}H+H\hat{\gamma}_{5}=0,\ \ \ 
\gamma_{5}D + D\hat{\gamma}_{5}=0,
\end{eqnarray}
and $\hat{\gamma}^{2}_{5}=1$ where
\begin{eqnarray}
\label{modifiedgamma5}
&&\hat{\gamma}_{5}=\gamma_{5}-2Hf(H^{2}),
\end{eqnarray}
which is the modified chiral matrix
~\cite{Niedermayer:1999bi,Narayanan:1998uu}. 
In this thesis we assume that the lattice Dirac operator 
satisfies the above general Ginsparg-Wilson relation and
has no doublers. We also assume that $D$ is local 
and has the smooth dependence 
with respect to link variables.

\subsection{Admissibility condition and
the index theorem}

The locality and smoothness have been established
for the overlap Dirac 
operator~\cite{Hernandez:1999et,Neuberger:2000pz}, 
if the plaquette variables satisfy 
the admissibility condition:
\begin{eqnarray}
\label{admiss}
&&\| 1-R[U_{\mu\nu}(x)]\| < \epsilon,\qquad 
\mbox{for all plaquettes},
\end{eqnarray} 
where $R$ is the representation of the gauge 
group~\footnote{Throughout this thesis, we assume that the
representation of the gauge group, $R$, is unitary.} 
and $\epsilon$ is a fixed positive 
number.\footnote{ As for $H$ satisfying 
the general Ginsparg-Wilson relation
with $f(H^2)=H^{2k}$ with a positive integer $k$,
the locality and the smoothness has been proved in the case of
free $H$~\cite{fujilocal}.}
We assume this condition also when we use Dirac operators 
satisfying the 
general Ginsparg-Wilson
relation~(\ref{generalGW}),\footnote{Note that
this restriction on link variables is 
irrelevant in the classical 
continuum limit.} since this restriction gives the topological
structure to the space of link variables~\cite{luschertopo,
stonetopo}.\footnote{In the case of $U(1)$ theory in finite 
volume, the topological 
structure of the space of admissible link variables has 
been investigated~\cite{Luscher:1999du}.} 
If there is no restriction on link 
variables, any link variables can be continuously
deformed to the trivial field $U_\mu(x)=1$ on the lattice and  
thus there is no topological structure.

Now we define the topological charge to 
establish the index theorem on the lattice.
For this purpose, we first discuss lattice chiral
symmetry~\cite{luscherchiral}. 
The fermion action which has the Dirac
operator satisfying the general Ginsparg-Wilson 
relation~(\ref{generalGW}) is invariant under 
the global transformation:
\begin{eqnarray}
&&\psi(x)\to \psi^\prime(x)=\psi(x)+i\epsilon^a T^a
\hat{\gamma}_5\psi(x),\nonumber\\
&& \bar{\psi}(x)\to \bar{\psi}^\prime(x)=\bar{\psi}(x)
+i\epsilon^a \bar{\psi}(x)\gamma_5 T^a,
\end{eqnarray}
on account of the relation~(\ref{varietyGW}) and where $T^a$ are  
flavor-matrices.
But the fermion
measure produces the Jacobian factor:
\begin{eqnarray}
&&{\cal D} \bar{\psi}^\prime{\cal D}\psi^\prime 
=J{\cal D}\bar{\psi}{\cal D}\psi,\nonumber\\
&& J=\mbox{exp}(-2i\epsilon\Tr(\Gamma_5T^a) ).
\end{eqnarray}
From this expression, we see that 
the fermion measure is invariant under 
the non-singlet flavor transformation, but not for the 
flavor-singlet 
transformation, which agrees with the continuum theory~\cite{fujim}. 
$\Tr \Gamma_5$ is regarded as the 
chiral anomaly on the lattice. Therefore we may define
the topological charge on the 
lattice as $Q=\Tr \Gamma_5$~\footnote{In fact, we can
show $\delta\left(\Tr \Gamma_5\right)=0$ under 
the local deformation of the link variables explicitly,
using the general Ginsparg-Wilson 
algebra~(\ref{generalGW}).}.
Now we can show the index theorem on the lattice by
examining the representation of the 
algebra~(\ref{generalGW})~\cite{Fujikawa:2000my}.
From the analysis in Appendix \ref{eigenh}, 
$\Tr \Gamma_5$ is written as 
follows($n=0$ stands for $\lambda_n =0$), 
\begin{eqnarray}
\label{latindex}
\Tr \Gamma_5 &\equiv& \sum_{n}
\left(\varphi_n, \Gamma_5\varphi_n\right)\nonumber\\
&=& \sum_{n=0}
\left(\varphi_n, \Gamma_5\varphi_n\right)
\nonumber\\
&=&\sum_{n=0}
\left(\varphi_n, \gamma_5\varphi_n\right)
=n_+ -n_-=\>\mbox{index}.
\end{eqnarray}
This is the index theorem on the 
lattice~\cite{Hasenfratz:1998ri}. 
Since $\Gamma_5$ depends smoothly on the link variable 
within the space of
admissible configurations~(\ref{admiss}), the integer $n_+-n_-$
is a constant in a connected component of the space of admissible
configurations; $n_+-n_-$ thus provides a 
topological characterization
of the gauge field 
configuration, i.e., the index~\footnote{In the classical 
continuum limit 
$\Tr \Gamma_5$ has been evaluated for the overlap Dirac
operator~\cite{kikuyama}-\cite{suzuano} 
and for the Dirac operators satisfying the general
Ginsparg-Wilson relation 
with $f(H^2)=H^{2k}$~\cite{fujichiral}. From their analysis,
the same index theorem as that of the continuum theory:
\begin{eqnarray}
\int d^4 x \frac{g^2}{32\pi^2}\tr\,\epsilon^{\mu\nu\rho\sigma}
F_{\mu\nu}F_{\rho\sigma}=n_+ - n_-
\end{eqnarray}
has been obtained in the classical continuum limit. }.

Therefore the admissibility condition~(\ref{admiss}) has important
roles in the formulation with the Ginsparg-Wilson relation.
The one role is that it ensures the locality
and smoothness of the lattice 
Dirac operator. Another role is that it 
gives rise to the topological structure in the space of
link variables.

\subsection{Gauge field action}
In this subsection let us mention 
the gauge field action in this formulation.
To impose the restriction on the space of link variables,
which is caused by the admissibility condition,
we may take a gauge field action which
is different from the standard Wilson 
action~(\ref{wilsonaction}).
For example,\footnote{ The other action 
has been proposed in the abelian chiral 
gauge theory~\cite{Luscher:1999du}:
\begin{eqnarray}
&&S_G=
\left\{
\begin{array}{ll}
\displaystyle
\frac{1}{g_0^2}a^4\sum_{x,\mu\neq\nu}
\frac{[F_{\mu\nu}(x)]^2}{1-[F_{\mu\nu}(x)]^2/\epsilon^2}   
&\quad \mbox{if} 
\>\| F_{\mu\nu}\|<\epsilon, \\
\infty &\quad \mbox{otherwise},
\end{array}
\right.
\end{eqnarray}
where $\epsilon$ is a constant in the 
range $0<\epsilon<\pi/3$ and $F_{\mu\nu}(x)$ is
defined through 
\begin{eqnarray}
&&F_{\mu\nu}(x)=\frac{1}{i}\mbox{ln}\,U_{\mu\nu}(x),
\qquad -\pi<F_{\mu\nu}(x)\le \pi. 
\end{eqnarray}
}
\begin{eqnarray}
\displaystyle
&&S_G=
\left\{
\begin{array}{ll}
\displaystyle
\frac{1}{g_0^2}a^4\sum_{x,\mu\neq\nu}
\frac{\Tr\left(1-U_{\mu\nu}(x)\right)}{1-
\| 1-U_{\mu\nu}(x)\|^2/\epsilon^2},
&\quad \mbox{if $\| 1-U_{\mu\nu}\|<\epsilon$}, \\
\infty &\quad \mbox{otherwise}. 
\end{array}\right.\nonumber\\
\end{eqnarray}
Note that $S_G$ is local and a smooth 
function of the link variables and satisfies gauge invariance. 
In particular, the 
functional integral can be constructed in the usual
way with the standard integration measure

\section{Lattice chiral gauge theory}
In this section, we analyze 
the case of pure chiral gauge theory without Higgs 
couplings. 
The extension to the case with Higgs couplings 
is straightforward and it will be discussed in 
Section~\ref{secyukawa}.

\subsection{Weyl fermion} 
To make the following discussion as general as possible, 
we introduce a
one-parameter family of lattice analog of $\gamma_5$:
\begin{equation}
\gamma_5^{(t)}={\gamma_5-tHf(H^2)\over\sqrt{1+
t(t-2)H^2f^2(H^2)}}.
\label{eq:twoxfour}
\end{equation}
Note that $\gamma_5^{(0)}=\gamma_5$ and, when $f(H^2)=1$,
$\gamma_5^{(2)}=\hat\gamma_5$ corresponds to the conventional 
modified
chiral matrix~(\ref{modifiedgamma5}). 
Note 
that $\gamma_5^{(1)}=\Gamma_5/\sqrt{\Gamma_5^2}$. The ``conjugate''
of $\gamma_5^{(t)}$ is defined by
\begin{equation}
   \overline\gamma_5^{(t)}=\gamma_5\gamma_5^{(2-t)}\gamma_5,
\label{eq:twoxseven}
\end{equation}
and they satisfy the following relations:
\begin{eqnarray}
   &&\gamma_5^{(t)\dagger}=\gamma_5^{(t)},\qquad
   \overline\gamma_5^{(t)\dagger}=\overline\gamma_5^{(t)},\qquad
   (\gamma_5^{(t)})^2=(\overline\gamma_5^{(t)})^2=1,
\nonumber\\
   &&\overline\gamma_5^{(t)}D+D\gamma_5^{(t)}=0.
\label{eq:twoxeight}
\end{eqnarray}
In view of the last relation in (\ref{eq:twoxeight}), 
we may introduce the chiral projection
operators by
\begin{equation}
   P_\pm^{(t)}={1\over2}(1\pm\gamma_5^{(t)}),\qquad
   \overline P_\pm^{(t)}={1\over2}(1\pm\overline\gamma_5^{(t)}),
\label{eq:twoxnine}
\end{equation}
so that
\begin{equation}
   \overline P_\pm^{(t)}D=DP_\mp^{(t)}.
\label{eq:twoxten}
\end{equation}
Then the left-handed Weyl fermion is defined 
by
\begin{equation}
   P_-^{(t)}\psi=\psi,\qquad\overline\psi\overline P_+^{(t)}
=\overline\psi.
\label{eq:twoxeleven}
\end{equation}
The kinetic term is then consistently decomposed according to the
chirality:
\begin{eqnarray}
D=\overline P_+^{(t)}DP_-^{(t)} + \overline P_-^{(t)}DP_+^{(t)}.
\end{eqnarray}

\subsection{Fermion integration measure}
To define the fermion integration measure, we introduce certain
orthonormal vectors in the constrained space,
\begin{eqnarray}
&&   P_-^{(t)}v_j(x)=v_j(x),\qquad
   \overline v_k\overline P_+^{(t)}(x)=\overline v_k(x),\nonumber\\
&& \left(v_j,v_k\right)=\delta_{jk},\qquad 
\left(\bar{v}_j^\dagger, \bar{v}_k^\dagger\right)=\delta_{jk},
\label{eq:twoxsixteen}
\end{eqnarray}
and expand the fields as
\begin{equation}
   \psi(x)=\sum_jv_j(x)c_j,\qquad
   \overline\psi(x)=\sum_k\overline c_k\overline v_k(x).
\label{eq:twoxseventeen}
\end{equation}
Then the (ideal) integration measure is defined by
\begin{equation}
   {\rm D}[\psi]{\rm D}[\overline\psi]
   =\prod_j{\rm d}c_j\prod_k{\rm d}\overline c_k.
\label{eq:twoxeighteen}
\end{equation}
The condition~(\ref{eq:twoxsixteen}) shows that the basis vectors may
depend on the link variable because the chiral projectors depend on it.
However, this conditions alone do not determine how the basis vectors
depend on the link variables. We first note the identity
\begin{eqnarray}
&&\delta_\xi v_j=P_-^{(t)}\delta_\xi v_j+
(1-P_-^{(t)})\delta_\xi v_j,
\end{eqnarray}
where $\delta_\xi v_j$ is the variation of basis under  
an infinitesimal variation of the
link variables,
\begin{equation}
   \delta_\xi U_\mu(x)=a\xi_\mu(x)U_\mu(x),\qquad \xi_\mu(x)
=\xi_\mu^a(x)T^a,
\label{eq:twoxtwentyoneone}
\end{equation}
with $\xi_\mu^a(x)$ being any real vector field 
and then note that the
constraint~(\ref{eq:twoxsixteen})
implies
\begin{eqnarray}
&&(1-P_-^{(t)})\delta_\xi v_j=\delta_\xi P_-^{(t)}v_j.
\label{projv}
\end{eqnarray}
Consequently, the second term of the variation $\delta_\xi v_j$, 
$(1-P_-^{(t)})\delta_\xi v_j$, is fixed uniquely by 
(\ref{projv}). However, the
component of~$\delta_\xi v_j$ residing in the constrained space,
$P_-^{(t)}\delta_\xi v_j$, is not determined by the constraint
(\ref{eq:twoxsixteen}). This fact means that the fermion measure
is not specified uniquely and there is an ambiguity
of the phase. One can see this by noting that when one changes
basis by unitary transformation matrices ${\cal Q}$ and 
 $\bar{{\cal Q}}$ as 
\begin{eqnarray}
\label{basistrf}
&&v_j^\prime(x)=\sum_l v_l(x){\cal Q}_{lj},
\qquad c_j^\prime(x)=\sum_{l}{\cal Q}^{-1}_{jl}c_l,\nonumber\\
&&\bar{v}_j^\prime(x)=\sum_l \bar{{\cal Q}}_{jl}\bar{v}_l(x),
\qquad \bar{c}_j^\prime=\sum_{l}\bar{c}_l\bar{{\cal Q}}^{-1}_{lj},
\end{eqnarray} 
the fermion measure changes as follows,
\begin{eqnarray}
&&{\rm D}[\psi]{\rm D}[\overline\psi]^{\lc v^\prime,\bar{v}^\prime\rc}
=\det {\cal Q}\det\bar{{\cal Q}}
{\rm D}[\psi]{\rm D}[\overline\psi]^{\lc v,\bar{v}\rc},
\end{eqnarray}
where ${\rm D}[\psi]{\rm D}[\overline\psi]^{\lc .., ..\rc}$ 
stands for the measure defined by the basis in the curly bracket 
and where $\det {\cal Q}$ and $\det\bar{{\cal Q}}$ are pure phase 
factors, which may depend on the link variables since
the basis vectors depend on the link variables.
Thus the fermion measure has the freedom of the phase which may
depend on the link variables. How to fix this phase will
be discussed later. For the time being we assume that 
some particular choice of the basis has been made. 
The above general setup, which is closely related to the overlap
formulation in refs.~\cite{Narayanan:1993wx,
Randjbar-Daemi:1995sq}, does
not specify the formulation uniquely, leaving the phase 
unspecified.

\subsubsection{Correlation functions and the effective action}
We have just defined the Weyl fermion and the fermion 
measure(except the freedom of 
the phase of the fermion integration measure).
Now we can describe the path-integral 
expression of the expectation value of
an operator ${\cal O}$~\cite{Luscher:1999du,Luscher:2000un}:
\begin{equation}
   \langle{\cal O}\rangle
   ={1\over{\cal Z}}\sum_M\int_M{\rm D}[U]\,e^{-S_G}
   {\cal N}_Me^{i\vartheta_M}\langle{\cal O}\rangle_{\rm F}^M,
\label{eq:twoxtwelve}
\end{equation}
where $M$ denotes the topological sector specified by the
admissibility~(\ref{admiss}).
The ``topological weight'', ${\cal N}_Me^{i\vartheta_M}$, 
with which the
topological sectors are summed, is not fixed within this 
formulation. In
each topological sector~$M$, the average with respect to the fermion
fields~$\langle{\cal O}\rangle_{\rm F}^M$ is defined by the generating
functional of fermion Green's functions
\begin{equation}
   Z_{\rm F}[U,\eta,\overline\eta;t]
   =\int{\rm D}[\psi]{\rm D}[\overline\psi]\,e^{-S_{\rm F}},
\label{eq:twoxfourteen}
\end{equation}
where we have written the dependence on the parameter~$t$ in
eq.~(\ref{eq:twoxfour}) 
explicitly~\footnote{The dependence on $t$ is
expected to disappear in the continuum limit, 
if the formulation is well-defined.} 
and
\begin{equation}
   S_{\rm F}=a^4\sum_x[\overline\psi(x)D\psi(x)-\overline\psi(x)
\eta(x)
   -\overline\eta(x)\psi(x)],
\label{eq:twoxfifteen}
\end{equation}
where $\eta(x)$ and $\bar{\eta}(x)$ are source fields. 
The fermion integration variables are subject to the chirality
constraint~(\ref{eq:twoxeleven}).

Now we can work out a few quantities in this formalism.
\begin{itemize}
 \item {\sf Fermion propagator}\\ 
If $D$ has not zero modes, we obtain
\begin{eqnarray}
\frac{\langle \psi(x)\bar{\psi}(y)\rangle}
{\langle 1\rangle_{\rm F}}&=&P_-^{(t)}\frac{1}{D}\bar{P}_+^{(t)}.
\end{eqnarray}
The correlation functions with arbitrary products of fermions
can be calculated by Wick's theorem.

\item {\sf Effective action}\\
In the vacuum sector the effective action without source fields
is given by
\begin{equation}
   \langle1\rangle_{\rm F}=\det M,\qquad
   M_{kj}=a^4\sum_x\overline v_k(x)Dv_j(x).
\label{eq:twoxtwentyfive}
\end{equation}
Its link variable variation is given by
\footnote{If one assumes that $v_j$ and~$\overline v_k$
change gauge covariantly under an infinitesimal 
gauge transformation,
$\delta_\xi v_j(x)=R(\omega(x))v_j(x)$
and~$\delta_\xi\overline v_k(x)=-\overline v_k(x)R(\omega(x))$, then
the effective action~(\ref{eq:twoxtwentyfive}) becomes invariant under
the gauge transformation. This assumption of the specific gauge
variation of basis vectors, however, modifies the physical
contents of the theory in general, because it does not reproduce the
gauge anomaly. We will see such an example of basis vectors in Sec.3.2.}
\begin{equation}
   \delta_\xi\ln\det M
   =\Tr(\delta_\xi DP_-^{(t)}D^{-1}\overline P_+^{(t)})
   -i{\mathfrak L}_\xi,
\label{eq:twoxtwentysix}
\end{equation}
where 
\begin{eqnarray}
\label{measureterm}
&&{\mathfrak L}_\xi[U;t]=i\sum_{j}\left(v_j,\delta_\xi v_j\right)
+i\sum_k(\delta_\xi\overline v_k^\dagger,\overline v_k^\dagger).
\end{eqnarray}
This term appears because
the basis vectors $\{v,\bar{v}\}$ depend on the link variables.
In addition, under the basis transformations~(\ref{basistrf}) 
this term transforms as follows,
\begin{eqnarray}
&&{\mathfrak L}_\xi^{\{v^\prime,\bar{v}^\prime\}}=
{\mathfrak L}_\xi^{\{v,\bar{v}\}}+i\delta_\xi\ln\,\det {\cal Q}
+i\delta_\xi\ln\,\det \bar{{\cal Q}}.
\end{eqnarray}
This transformation law means that ${\mathfrak L}_\xi[U;t]$ 
is a quantity associated with the fermion measure and 
we call this term the ``measure term''.
Since the measure term is purely imaginary, 
it contributes only to the
imaginary part of eq.~(\ref{eq:twoxtwentysix}):
\begin{eqnarray}
   &&{1\over2}[\Tr(\delta_\xi DP_-^{(t)}D^{-1}\overline P_+^{(t)})
   -\Tr(\delta_\xi DP_-^{(t)}D^{-1}\overline P_+^{(t)})^*]
\nonumber\\
   &&={1\over2}[
   \Tr(\delta_\xi DP_-^{(t)}D^{-1})
   -\Tr(\delta_\xi DP_+^{(2-t)}D^{-1})]
\nonumber\\
   &&=-{1\over4}\Tr\delta_\xi D(\gamma_5^{(t)}
   +\gamma_5^{(2-t)})D^{-1},
\label{eq:twoxtwentyseven}
\end{eqnarray}
where we have used eq. (\ref{ghermicity}) 
and~eq.~(\ref{eq:twoxseven}). The measure term is 
then chosen to improve
the imaginary part of the effective action (see, for example,
ref.~\cite{Suzuki:1999qw}).

The measure term can be rewritten, when we take the 
linearity of the measure term
on $\xi_\mu(x)$ into account, as follows,
\begin{eqnarray}
&&{\mathfrak L}_\xi=a^4\sum_x \xi_\mu^a(x)j_\mu^a(x).
\end{eqnarray}
with a current $j_\mu^a(x)$. In the following subsection this
current associated with the measure plays an important role.
\end{itemize}

\subsection{L\"{u}scher's reconstruction 
theorem}\label{luschertheorem}
Let us return to the problem of how to fix the phase of 
the fermion measure which has been 
unspecified. We should find a basis in
eq.~(\ref{eq:twoxeighteen}) or the associated measure term 
for which the
gauge invariance (assuming the fermion multiplet is 
anomaly-free) and
the locality are ensured for finite lattice spacings in this 
formulation. 
L\"{u}scher has showed that chiral gauge theories with 
anomaly-free
multiplets of Weyl fermions\footnote{Anomaly-free 
fermion multiplets means that the 
following trace over gauge group matrices in 
the representation $R$ vanish:
\begin{eqnarray}
\label{anomalyfun}
d_R^{abc}=2i\tr\left[T^a\lc T^b,T^c\rc\right]=0.
\end{eqnarray}
} can be put on the lattice without
violating the fundamental principles if several  
conditions on the current $j_\mu(x)$ associated with the measure
term are satisfied, which is called
``L\"{u}scher's reconstruction theorem''~\cite{Luscher:1999du,
Luscher:2000un}.
\begin{flushleft}
{\bf L\"{u}scher's reconstruction theorem}
\end{flushleft}
The formulation of the chiral gauge theory 
with anomaly-free fermions on the lattice is local and smooth on 
the link variables and has gauge invariance,
if the current $j_\mu(x)$ satisfies the following 
conditions.\footnote{Then the ambiguity of the fermion 
measure is reduced to a constant phase which depends only on  
topological sectors.}  
\begin{description}
\item[i)locality and smoothness]\hspace{2ex}
$j_\mu(x)$ is local and is 
smoothly dependent on the link variables.

\item[ii)gauge invariance]\hspace{2ex}
$j_\mu(x)$ is gauge-covariant and satisfies 
the ``anomalous conservation law'':\footnote{Covariant 
derivatives in these expressions are defined by
\begin{eqnarray}
   &&\nabla_\mu^*j_\mu(x)
   ={1\over a}[j_\mu(x)-U_\mu(x-a\hat\mu)^{-1}j_\mu(x-a\hat\mu)
   U_\mu(x-a\hat\mu)].
\end{eqnarray}
}
\begin{eqnarray}
(\nabla_\mu^*j_\mu)^a(x)&=&{\cal A}^a(x)\nonumber\\
{\cal A}^a(x)&=&-{i\over2}\tr R(T^a)
   (\gamma_5^{(t)}+\gamma_5^{(2-t)})(x,x).
\label{anomalous}
\end{eqnarray}

\item[iii)integrability]\hspace{2ex}
$j_\mu(x)$ satisfies 
the ``integrability condition'':
Along all closed curves in the space 
of the link variables $U^\tau_\mu(x)$  
($0\leq\tau\leq1$), Wilson lines which are made of $j_\mu(x)$
satisfy the following equation:
\begin{eqnarray}
W&=&\exp\biggl(i\int_0^1\rmd\tau\,{\mathfrak L}_\xi\biggr),\quad
   a\xi_\mu(x)=\partial_\tau U^\tau_\mu(x)U^\tau_\mu(x)^{-1},
\nonumber\\
  &=&\det(1-P_0^{(t)}+P_0^{(t)}Q_1^{(t)})
   \det\nolimits(1-P_0^{(2-t)}+P_0^{(2-t)}Q_1^{(2-t)}),\nonumber\\
\label{integral}
\end{eqnarray}
where $P_\tau^{(t)}=P_-^{(t)}|_{U=U_\tau}$
and~$\overline P_\tau^{(t)}=\overline P_+^{(t)}|_{U=U_\tau}$. The
unitary operator~$Q_\tau^{(t)}$ in this expression is defined by
\begin{equation}
   \partial_\tau Q_\tau^{(t)}
   ={1\over4}[\partial_\tau\gamma_\tau^{(t)},\gamma_\tau^{(t)}]
   Q_\tau^{(t)},\qquad Q_0^{(t)}=1,
\label{eq:fivexfour}
\end{equation}
where $\gamma_\tau^{(t)}=\gamma_5^{(t)}|_{U=U_\tau}$. 
\end{description}
The current $j_\mu(x)$ which has above properties is 
called an ``ideal current''.\footnote{We note that 
the ideal current exists only for anomaly-free multiplets.} 
The next crucial task is 
to find out an ideal current.  
For $\U(1)$
theories, such a current was constructed by
L\"uscher~\cite{Luscher:1999du}. 
For electroweak $\SU(2)\times\U(1)$ theory also, such a current 
has been constructed 
on the infinite volume lattice~\cite{Kikukawa:2001kd}.\footnote{
See~\cite{kikunakasuzu} on an attempt to construct on 
the finite volume lattice.} 
For general 
non-abelian theories, the
measure from an ideal current has been 
constructed only in perturbation
theory~\cite{Luscher:2000zd,Suzuki:2000ii},  
but the 
existence of such an
ideal fermion measure in non-perturbative level has not 
been established yet(except for the real representation 
of the gauge group~\cite{Suzuki:2000ku}).\footnote{
The Witten anomaly on the lattice also has been discussed and  
it has been shown that there exists no 
ideal current $j_\mu(x)$ in the chiral gauge theory with 
odd number of Weyl fermion multiplets. 
} 
In such a case it seems necessary to clarify the topological 
structure of the space of the admissible link variables but
this task remains to be performed.\footnote{As one of attempts 
toward this challenging task, see ref.\cite{adamstopo}.}

We will discuss more details of the contents of 
the above theorem in the following subsections. 

\subsubsection{Locality and smoothness}
We first require that the expectation 
value of the products of fermionic and link 
variables (\ref{eq:twoxtwelve}) should be 
smooth functions of the link variable. 
This requirement is necessary to derive the field equation
discussed below. 

The locality is one of the most important properties in  
lattice theory. The universality in the continuum limit
depends on this property. To ensure the locality in this formulation,
it is necessary to require the locality of field equation.
The field equation is obtained by calculating the change
of (\ref{eq:twoxtwelve}) under the variations of the link 
variables~(\ref{eq:twoxtwentyoneone}). When ${\cal O}$ does not 
contain fermion fields,\footnote{When more general ${\cal O}$ 
is considered, the field equation can not be derived easily. 
Because
all sectors with non-zero index contribute to this equation.} 
we obtain 
\begin{eqnarray}
&&\langle \{ \delta_\xi S_G + \sum_{x}\bar{\psi}(x)\delta_\xi 
D
\psi(x) +i{\mathfrak L}_\xi\} {\cal O}\rangle
=\langle \delta_\xi {\cal O}\rangle.
\end{eqnarray}    
In this equation the first two terms are clearly local owing 
to the locality of the action. We thus need to require that 
${\mathfrak L}_\xi$, 
therefore, $j_\mu(x)$ should be local functions 
of the link variable to ensure the locality of field equation
, as stated in {\bf i)} in above theorem.

\subsubsection{Gauge invariance}
The gauge invariance also is an important property which has to
be required in this formulation. In particular, the fermion 
effective action should be gauge-invariant. 
Now the gauge transformation is given 
by~(\ref{eq:twoxtwentyoneone}) 
with 
\begin{eqnarray}
\xi_\mu(x)&=&-\nabla_\mu \omega(x)\nonumber\\
&=&\-{1\over a}[U_\mu(x)\omega(x+a\hat\mu)
U_\mu(x)^{-1}-\omega(x)].
\end{eqnarray}
Noting the transformation property of the Dirac operator:
\begin{eqnarray}
\delta_\xi D =\lc R(w), D\rc,
\end{eqnarray}
the gauge variation of the expectation value is
written as follows,
\begin{eqnarray}
   &&\delta_\xi\langle{\cal O}\rangle_{\rm F}
   =\langle\delta_\xi{\cal O}\rangle_{\rm F}
   +ia^4\sum_x\omega^a(x)[{\cal A}^a(x)-(\nabla_\mu^*j_\mu)^a(x)]
   \langle{\cal O}\rangle_{\rm F},
\nonumber\\
   &&{\cal A}^a(x)=-{i\over2}\tr R(T^a)
   (\gamma_5^{(t)}+\gamma_5^{(2-t)})(x,x),
\label{eq:fivexsix}
\end{eqnarray}
where ${\cal A}^a(x)$ can be regarded as the covariant 
anomaly on the lattice. \footnote{In fact, 
we can show explicitly that ${\cal A}^a(x)$ corresponds to  
the covariant anomaly in the classical continuum limit  
in the case of $f(H^2)=1$~\cite{kikuyama}-\cite{suzuano} 
and $f(H^2)=H^{2k}$~\cite{fujichiral}. Then the measure 
term may be regarded as the Bardeen-Zumino term, which 
relates the covariant anomaly to the consistent anomaly and 
is local~\cite{bz}.}
Hence, for the gauge invariant formulation, 
the current $j_\mu(x)$ should satisfy the
``anomalous conservation law'':
\begin{equation}
   (\nabla_\mu^*j_\mu)^a(x)={\cal A}^a(x),
\label{eq:fivexeight}
\end{equation}
and it should be 
also gauge-covariant because of the gauge invariance 
of the effective action and (\ref{eq:twoxtwentysix}), 
as stated in {\bf ii)}.

\subsubsection{Integrability condition}
To obtain the fermion measure from a given current $j_\mu(x)$,  
we  need the integrability of a current. 
Here, 
we show that the fermion measure can be constructed from 
$j_\mu(x)$ which satisfies a certain integrability condition.

To derive the integrability condition,   
we consider the change of the phase of the fermion effective 
action under the smooth curves in the space of link variable:
\begin{eqnarray}
U^\tau_\mu(x),\qquad 0\le\tau\le 1.
\end{eqnarray}
Then the change of the phase of the effective action is represented
as Wilson line:
\begin{eqnarray}
\label{Wilsonline}
W&=&\exp\biggl(i\int_0^1\rmd\tau\,{\mathfrak L}_\xi\biggr),\quad
   a\xi_\mu(x)=\partial_\tau U^\tau_\mu(x)U^\tau_\mu(x)^{-1}.
\end{eqnarray}
We first calculate the Wilson line with 
a basis $\lc v,\bar{v}\rc$ along a closed curve, 
assuming that this basis is smoothly 
defined
on the space of admissible link variables.
To describe this we introduce the projectors
\begin{eqnarray}
P_\tau^{(t)}=P_-^{(t)}|_{U=U^\tau},\qquad
\overline P_\tau^{(t)}=\overline P_+^{(t)}|_{U=U^\tau},
\end{eqnarray}
and define the unitary operators $Q_\tau$ through
\begin{eqnarray}
  \partial_\tau Q_\tau^{(t)}
   =[\partial_\tau P_\tau^{(t)},P_\tau^{(t)}]
   Q_\tau^{(t)},\qquad Q_0^{(t)}=1,
\label{eq:fivexfour}
\end{eqnarray}
From this, we have (See Appendix~\ref{computations})
\begin{equation}
   P_\tau^{(t)}=Q_\tau^{(t)}P_0^{(t)}Q_\tau^{(t)\dagger},\qquad
   \overline P_\tau^{(t)}
   =\gamma_5Q_\tau^{(2-t)}\gamma_5\overline P_0^{(t)}
   (\gamma_5Q_\tau^{(2-t)}\gamma_5)^\dagger.
\label{eq:fivexfive}
\end{equation}
Therefore $Q_\tau$ is the transporter of $P_\tau$ along the curve.
After some calculations(Appendix~\ref{computations}), 
the Wilson line with the smooth basis
is obtained as follows, 
\begin{eqnarray}
 W&=&\det(1-P_0^{(t)}+P_0^{(t)}Q_1^{(t)})
   \det\nolimits^{-1}(1-\overline P_0^{(t)}
   +\overline P_0^{(t)}\gamma_5Q_1^{(2-t)}\gamma_5)
\nonumber\\
   &=&\det(1-P_0^{(t)}+P_0^{(t)}Q_1^{(t)})
   \det\nolimits(1-P_0^{(2-t)}+P_0^{(2-t)}Q_1^{(2-t)}),
\label{eq:fivexthree}
\end{eqnarray} 
along all closed curves, for which $U^1=U^0$. 

\begin{figure}[htbp]
\begin{flushright}
\includegraphics{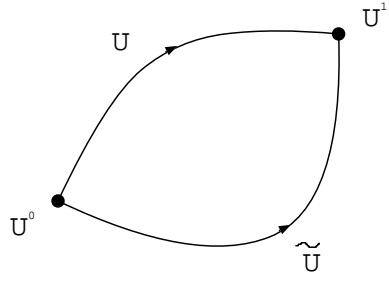}
\caption{Two curves in the space of link variables}
\label{fig.curve}
\end{flushright}
\end{figure}

\begin{figure}[htbp]
\begin{center}
\includegraphics{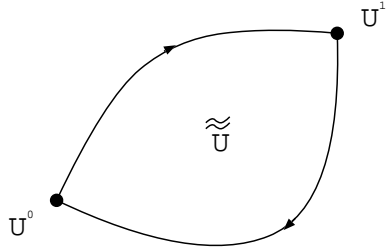}
\caption{Closed curve in the space of link variables}
\label{closedcurve}
\end{center}
\end{figure}

Now let $j_\mu(x)$ be the smooth function of the admissible 
link variables. 
That $j_\mu(x)$ should satisfy the 
above equation~(\ref{eq:fivexthree}) is clearly the necessary 
condition to obtain the fermion measure 
from the current $j_\mu(x)$. 
Moreover, this equation is also the sufficient condition and 
we can now prove the L\"{u}scher's 
reconstruction theorem as follows.\\
\\
{\bf Proof of the reconstruction theorem}\\
Now we assume that we have found an ideal current 
$j_\mu(x)$. 
The proof
is given by an explicit construction of 
the basis. We may set the fermionic
basis vectors for the gauge field $U$ as
\begin{equation}
   v_j=\cases{Q_1^{(t)}u_1W^{-1},&for $j=1$,\cr
              Q_1^{(t)}u_j,&otherwise,\cr}\qquad
   \overline v_k=\overline u_k(\gamma_5Q_1^{(2-t)}\gamma_5)^\dagger,
\label{eq:fivexnine}
\end{equation}
where $u_j$ and~$\overline u_k$ are fixed bases defined for the
reference configuration~$U^0$. The vectors are then transported by the
operators~$Q_\tau^{(t)}$ and~$Q_\tau^{(2-t)}$ along a certain curve
connecting $U^0$ and~$U=U^1$. The Wilson line~$W$ is also defined 
along
this curve. The above basis depends on the curve chosen, but the
associated measure does not, because of~eq.~(\ref{eq:fivexthree}).
We can show this as follows. We first consider the two basis, 
$\lc v,\bar{v}\rc$ and $\lc \tilde{v},\tilde{\bar{v}}\rc$, which 
are defined along two curves, $U^\tau$ and $\tilde{U}^\tau$,
which have the same start $U^0$ and end $U=U^1$, respectively~
(\ref{fig.curve}). 
Explicitly, $\lc v,\bar{v}\rc$ are given 
by eq.(\ref{eq:fivexnine}) and $\lc \tilde{v},\tilde{\bar{v}}\rc$ 
are given by
\begin{eqnarray}
  \tilde{v}_j=\cases{\tilde{Q}_1^{(t)}u_1
\tilde{W}^{-1},&for $j=1$,\cr
              \tilde{Q}_1^{(t)}u_j,&otherwise,\cr}\qquad
   \tilde{\overline v}_k=\overline u_k(\gamma_5\tilde{Q}_1^{(2-t)}
\gamma_5)^\dagger,
\end{eqnarray}  
where $\tilde{W}$ is constructed from $\tilde{Q}_1^{(t)}$ and 
$\tilde{Q}_1^{(2-t)}$ corresponding to the curve 
$\tilde{U}$. Now the change of the two fermion measures which 
are constructed from two bases is given by
\begin{eqnarray}
{\rm D}[\psi]{\rm D}[\overline\psi]^{\lc v,\bar{v}\rc}
&=&\det\lc (\tilde{v}_j, v_k)\rc\det\lc (\bar{v}_j^\dagger,
\tilde{\bar{v}}_k^\dagger)\rc
{\rm D}[\psi]{\rm D}[\overline\psi]^{\lc 
\tilde{v},\tilde{\bar{v}}\rc}\nonumber\\
&=&\left(\tilde{W}^{-1}W\right)^{-1}\det\lc (u_j,\tilde{Q}_1^{(t)
\dagger}Q_1^{(t)}u_k)\rc \nonumber\\
&&\times\det \lc (\bar{u}_j^\dagger,
(\gamma_5\tilde{Q}_1^{(2-t)\dagger}Q_1^{(2-t)}\gamma_5)^\dagger
\bar{u}_k^\dagger)\rc{\rm D}[\psi]{\rm D}[\overline\psi]^{\lc 
\tilde{v},\tilde{\bar{v}}\rc}.\nonumber\\
\end{eqnarray}
When we define a Wilson line $\tilde{\tilde{W}}$ and 
$\tilde{\tilde{Q}}_1^{(t)}$  
for a closed curve $\tilde{\tilde{U}}^\tau$ constructed 
from $U$ and $\tilde{U}^{-1}$ in fig.(\ref{closedcurve}), 
we have 
\begin{eqnarray}
&&\tilde{\tilde{W}}=\tilde{W}^{-1}W,\qquad \tilde{\tilde{Q}}_1^{(t)}
=\tilde{Q}_1^{(t)\dagger}Q_1^{(t)},  
\end{eqnarray}
from the definition of Wilson line~(\ref{Wilsonline}) 
and the transporter of projection 
operators~(\ref{eq:fivexfour}). 
Then the change of two measures is given by
\begin{eqnarray}
&&\det\lc (\tilde{v}_j, v_k)\rc\det\lc (\bar{v}_j^\dagger,
\tilde{\bar{v}}_k^\dagger)\rc\nonumber\\
&&\qquad =\det\nolimits^{-1}
(1-P_0^{(t)}+P_0^{(t)}\tilde{\tilde{Q}}_1^{(t)})
   \det\nolimits^{-1}(1-P_0^{(2-t)}+P_0^{(2-t)}
\tilde{\tilde{Q}}_1^{(2-t)})\nonumber\\
&&\qquad\quad\times 
\det\nolimits(1-P_0^{(t)}+P_0^{(t)}\tilde{\tilde{Q}}_1^{(t)}) 
\det\nolimits(1-P_{+0}^{(2-t)}+P_{+0}^{(2-t)}
\tilde{\tilde{Q}}_1^{(2-t)\dagger})\nonumber\\
&&\qquad =1.
\end{eqnarray}
Thus the measure is path-independent. 

We next compute the measure term~(\ref{measureterm}) 
for the above 
basis~(\ref{eq:fivexnine}). Under 
$\left.\delta_\xi U_\mu^\tau(x)=
\partial_\tau U_\mu^\tau(x)\right|_{\tau=1}$, the variations 
of bases are given by
\begin{eqnarray}
&&\delta_\xi v_j=\cases{\lb \partial_\tau P_\tau^{(t)}|_{\tau=1}, 
P_1^{(t)}\rb Q_1^{(t)} u_1 W^{-1} \cr
\qquad -iQ_1^{(t)}u_1\left(a^4\sum_{y}\xi_\mu^a(y) j_\mu^a(y)\right)
W^{-1}, &\mbox{for $j=1$},  \cr\cr
\lb \partial_\tau P_\tau^{(t)}|_{\tau=1}, 
P_1^{(t)}\rb Q_1^{(t)} u_j, & \mbox{otherwise}, \cr}
\end{eqnarray}
and 
\begin{eqnarray}
&&\delta_\xi \bar{v}_k=
\bar{u}_k(\gamma_5 Q_1^{(2-t)}\gamma_5)^\dagger
\left(\lb \partial_\tau \bar{P}_\tau^{(t)}|_{\tau=1}, 
\bar{P}_1^{(t)}\rb\right)^\dagger,
\end{eqnarray} 
where we used 
\begin{eqnarray}
  \partial_\tau (\gamma_5 Q_\tau^{(2-t)}\gamma_5)
   =[\partial_\tau \bar{P}_\tau^{(t)},\bar{P}_\tau^{(t)}]
   \gamma_5 Q_\tau^{(2-t)}\gamma_5,
\end{eqnarray}
which is derived from~(\ref{eq:fivexfour}). 
When we substitute these expressions for the variations of 
the bases in~(\ref{measureterm}), we obtain 
\begin{eqnarray}
&&{\mathfrak L}_\xi^{\{v,\bar{v}\}}
=a^4\sum_{x}\xi_\mu^a(x) j_\mu^a(x),
\end{eqnarray}
utilizing that $u_j(x)$ are orthonormal vectors. Since 
the ideal current $j_\mu^a(x)$ 
satisfies conditions (i) and (ii), the formulation 
is local and gauge invariant. 
\vspace{5ex}\\
In this chapter we have discussed the construction of lattice 
chiral gauge theory with the exact gauge invariance.  
In the next chapter, we discuss the
effect of CP breaking in this formulation by simply assuming the
existence of such an ideal measure.\footnote{Note however that our
analysis is relevant for a manifestly gauge invariant perturbation
theory~\cite{Luscher:2000zd} based on this formulation.} 
The properties of an ideal basis, 
which is summarized in L\"{u}scher's 
reconstruction theorem, 
play a crucial role in our following analysis.

\chapter{CP breaking in lattice chiral gauge theories}
In this chapter we discuss CP breaking in L\"{u}scher's formulation 
of lattice chiral gauge theory. We first show that the chiral 
fermion action with a local and doubler-free Ginsparg-Wilson 
operator has no CP symmetry under rather mild assumptions 
for projection operators~\cite{ffis}. 
We next calculate the fermion generating 
functional and investigate where 
the effects of CP breaking 
appear in this formulation with projection 
operators (\ref{eq:twoxnine})\cite{fiss}.

\section{Lattice CP problem in chiral fermion action }
In this section we discuss the breaking of CP symmetry in chiral 
fermion action. We first define C, P, CP transformations on 
the lattice. Next, we generalize the analysis of 
Hasenfratz~\cite{Hasenfratz:2001bz} 
and show that CP breaking is an inherent 
feature of this formulation.

\subsection{C, P and CP symmetry on the lattice} 
In this subsection we discuss C, P and CP transformations 
in this order. 
From now on, we assume that the vector-like theory with a  
Dirac operator which satisfy the 
general Ginsparg-Wilson relation~(\ref{generalGW}) is invariant 
under the standard C, P, and CP transformations.
The Wilson fermion action~(\ref{gaugeWilsonfer})
has these discrete symmetries as shown below. 
Therefore, 
the above assumption is correct for the action with 
the Dirac operators which 
are constructed out of the Wilson-Dirac operators, for instance, 
for the action of the overlap Dirac operator~(\ref{overlap}). 
We will also show 
that this assumption is consistent with the general 
Ginsparg-Wilson relation~(\ref{generalGW}). 

\subsubsection{Charge conjugation}
We denote the charge conjugation by C. 
The charge conjugation is defined by~\footnote{The following 
charge conjugation and 
parity transformation on the lattice are defined by an analogy with 
those in continuum theory(and in Minkowski space). 
See, for example, the book~\cite{BD} 
for the charge conjugation and the parity operation in continuum 
theory.} 
\begin{eqnarray}
\label{chargeconj}
   &&\psi(x)\to-C^{-1}\overline\psi^T(x),\qquad
   \overline\psi(x)\to\psi^T(x)C,
\nonumber\\
   &&U_\mu(x)\to U^{\rm C}_\mu(x)=U_\mu(x)^*,
\end{eqnarray}
where the charge conjugation matrix~$C=\gamma_2\gamma_4$ satisfies
\begin{equation}
   C^\dagger C=1,\qquad C^T=-C,\qquad C\gamma_\mu C^{-1}
=-\gamma_\mu^T,
   \qquad C\gamma_5C^{-1}=\gamma_5^T.
\end{equation}
Now the Wilson fermion action~(\ref{gaugeWilsonfer}) transforms 
under 
this charge conjugation as follows,  
\begin{eqnarray}
\label{ctrfgaugeWilsonfer}
&\to& a^4\sum_{x,\mu}\frac{1}{2a}\psi^T(x)\lb
-C\gamma_\mu C^{-1}\left(U_\mu^*(x)\bar{\psi}^T(x+a\hat{\mu})
\right.\right.\nonumber\\
&&\left.\qquad\qquad\qquad\qquad\qquad -U_\mu^T(x-a\hat{\mu})
\bar{\psi}^T(x-a\hat{\mu})\right)
\nonumber\\
&&\left.\quad -r\left(2\bar{\psi}^T(x)-U_\mu^*(x)
\bar{\psi}^T(x+a\hat{\mu}) 
-U_\mu^T(x-a\hat{\mu})\bar{\psi}^T(x-a\hat{\mu})\right)\rb
\nonumber\\
&&\quad -\frac{m_0}{a}\psi^T(x)\bar{\psi}^T(x)\nonumber\\
&&=(S_F^{(W)})^T=S_F^{(W)}.
\end{eqnarray}
Therefore, the Wilson fermion action is invariant under 
the charge conjugation. 
 
This invariance of the action can be represented in terms of 
the kernel of the Wilson-Dirac operator $D_W(x,y)$ 
as follows,
\begin{eqnarray}
D_W(U^C)(x,y)=C^{-1} D_W(U^C)(x,y)^T C,
\end{eqnarray} 
where the transpose operation~$T$ acts not only 
on the matrices involved
but also on the arguments as $(x,y)\to(y,x)$.
The kernel of 
the Dirac operator similarly transforms as
\begin{equation}
\label{diraccharge}
   D(U^{\rm C})(x,y)=C^{-1}D(U)(x,y)^TC. 
\end{equation}
Moreover we then have
\begin{eqnarray}
 && H(U^{\rm C})(x,y)=C^{-1}(\gamma_5 H(U)(x,y)\gamma_5)^T C,
\nonumber\\
&&  H(U^{\rm C})(x,y)^2=C^{-1}(H(U)(x,y)^2)^T C.
\end{eqnarray}
From this transformation laws, we see that the charge 
conjugation~(\ref{chargeconj}) is 
consistent with the general Ginsparg-Wilson 
relation~(\ref{generalGW}).

\subsubsection{Parity transformation}
We denote P the parity transformation by P:
\begin{eqnarray}
&&x=(x_i,x_4)\stackrel{P}{\to} \bar{x}=(-x_i,x_4),
\end{eqnarray}
where $i=1,2,3$. The parity transformation for the fields
is defined by
\begin{eqnarray}
   &&\psi(x)\to\gamma_4\psi(\bar x),\qquad
   \overline\psi(x)\to\overline\psi(\bar x)\gamma_4,
\nonumber\\
   &&U_\mu(x)\to U^{\rm P}_\mu(x)=\cases{
   U_i(\bar x-a\hat i)^{-1},&for $\mu=i$,\cr
   U_4(\bar x),& for $\mu=4$,\cr}.
\end{eqnarray}
Under this parity transformation, the Wilson fermion action 
transforms as
\begin{eqnarray}
&&\to  a^4\sum_{x,\mu}\frac{1}{2a}\bar{\psi}(\bar{x})
\lb
\gamma_4\gamma_i\gamma_4 \left(U_i^\dagger(\bar{x}-a\hat{i})
\psi(\bar{x}-a\hat{i})
-U_i^\dagger(\bar{x})
\psi(\bar{x}+a\hat{i})\right)
\right.\nonumber\\
&&\qquad\qquad +\gamma_4
\left(U_4(\bar{x})
\psi(\bar{x}+a\hat{4})-U_4^\dagger(\bar{x})
\psi(\bar{x}-a\hat{4})\right)\nonumber\\
&&\qquad\qquad 
+r\left(2\psi(\bar{x})-U_i^\dagger(\bar{x}-a\hat{i})
\psi(\bar{x}-a\hat{i}) 
-U_i(\bar{x})
\psi(\bar{x}+a\hat{i})\right)
\nonumber\\
&&\qquad\qquad\left.
+r\left(-U_4(\bar{x})
\psi(\bar{x}+a\hat{4}) 
-U_4^\dagger(\bar{x}-a\hat{4})
\psi(\bar{x}-a\hat{4})\right)\rb
\nonumber\\
&&\qquad\qquad 
+\frac{m_0}{a}\bar{\psi}(\bar{x})\psi(\bar{x})\nonumber\\
&&=S_F^{(W)}, 
\end{eqnarray}
where we used $\sum_{x}=\sum_{\bar{x}}$. 
Therefore, the Wilson fermion action satisfies P symmetry 
and the kernel of the Dirac operator transforms as follows,
\begin{equation}
\label{diracparity}   
D(U^{\rm P})(x,y)=\gamma_4D(U)(\bar{x},\bar{y})\gamma_4.
\end{equation}
under the parity transformation.\footnote{We can show 
eq. (\ref{diraccharge}) and (\ref{diracparity}) also for the Dirac 
operators satisfying the Ginsparg-Wilson relation 
with $f(H^2)=H^{2k}$~(\ref{gwo2}).} 
Further, we have
\begin{eqnarray}
&&H(U^{\rm P})(x,y)=-\gamma_4H(U)(\bar{x},\bar{y})\gamma_4,
\nonumber\\
&&H(U^{\rm P})(x,y)^2=\gamma_4H(U)(\bar{x},\bar{y})^2\gamma_4.
\end{eqnarray}
Therefore this parity transformation is consistent with the general 
Ginsparg-Wilson relation~(\ref{generalGW}).

\subsubsection{CP transformation}
CP transformation interchanges the fermion and anti-fermion.
Combining the charge conjugation with the parity transformation, 
CP transformation is written as follows,
\begin{eqnarray}
\label{cptrf}   
&&\psi(x)\to-W^{-1}\overline\psi^T(\bar x),\qquad
   \overline\psi(x)\to\psi^T(\bar x)W
\nonumber\\
   &&U_\mu(x)\to U^{\rm CP}_\mu(x)=\cases{
   {U_i(\bar x-a\hat i)^{-1}}^*,&for $\mu=i$,\cr
   U_4(\bar x)^*,& for $\mu=4$,\cr}
\end{eqnarray}
where
\begin{eqnarray}
   &&W=\gamma_2,\qquad W^\dagger W=1,
\nonumber\\
   &&W\gamma_\mu W^{-1}
   =\cases{\gamma_i^T,&for $\mu=i$,\cr
   -\gamma_4^T,& for $\mu=4$,\cr}\qquad
   W\gamma_5 W^{-1}=-\gamma_5^T,
\end{eqnarray}
and thus CP acts on the plaquette variables $U_{\mu\nu}(x)$ as
($\varphi_{ij}(\bar x-a\hat i-a\hat j)%
=U_j(\bar x-a\hat i-a\hat j)^*U_i(\bar x-a\hat i)^*$)
\begin{eqnarray}
   &&U_{ij}(x)
\nonumber\\
   &&\to U^{\rm CP}_{ij}(x)
   =\varphi_{ij}(\bar x-a\hat i-a\hat j)^{-1}
   U_{ij}(\bar x-a\hat i-a\hat j)^*
   \varphi_{ij}(\bar x-a\hat i-a\hat j),
\nonumber\\
   &&U_{i4}(x)\to U^{\rm CP}_{i4}(x)
   ={U_i(\bar x-a\hat i)^*}^{-1}
   {U_{i4}(\bar x-a\hat i)^{-1}}^*U_i(\bar x-a\hat i)^*.
\end{eqnarray}
Of course, we can show explicitly that the Wilson fermion action 
is invariant under 
this CP transformation and we also have 
\begin{equation}
   D(U^{\rm CP})(x,y)=W^{-1}D(U)(\bar x,\bar y)^TW.
\end{equation}
See Appendix \ref{conventions} for the CP transformation 
properties of various operators.

\subsection{Lattice CP problem}\label{cppro}

It has been pointed out that the CP symmetry, the 
fundamental
discrete symmetry in chiral gauge theories, is not manifestly
implemented in the L\"{u}scher's  formulation of lattice 
chiral gauge theory. The basic observation 
related to this
effect is as follows~\cite{Hasenfratz:2001bz}: In the 
formulation, the
chirality is imposed through~\cite{Narayanan:1998uu,
Niedermayer:1999bi}
\begin{equation}
   P_-\psi=\psi,\qquad\overline\psi\overline P_+=\overline\psi,
\label{eq:onexone}
\end{equation}
where $\overline P_\pm=(1\pm\gamma_5)/2$ and
$P_\pm=(1\pm\hat\gamma_5)/2$ with 
$\hat\gamma_5=\gamma_5(1-2aD)$.  
However,
since the above condition is not symmetric in the fermion and
anti-fermion and since CP exchanges these two, 
CP symmetry is explicitly
broken. In fact, the fermion action in the case of pure 
chiral gauge
theory without Higgs couplings changes under CP 
transformation~(\ref{cptrf}) as follows
\begin{equation}
   S_{\rm F}=a^4\sum_x\overline\psi(x)\overline P_+D P_-\psi(x)
   \to S_{\rm F}=a^4\sum_x\overline\psi(x)\gamma_5P_+
   \gamma_5D\overline P_-\psi(x),
\end{equation}
and this causes the change in the propagator 
\begin{eqnarray}
  && {\langle\psi(x)\overline\psi(y)\rangle_{\rm F}
   \over\langle1\rangle_{\rm F}}
  =P_-{1\over D}\overline P_+(x,y),
\nonumber\\
   &&\to{\langle\psi(x)\overline\psi(y)\rangle_{\rm F}
   \over\langle1\rangle_{\rm F}}
   =\overline P_-{1\over D}\gamma_5 P_+\gamma_5(x,y)\nonumber\\
&&\qquad\qquad
   =P_-{1\over D}\overline P_+(x,y)
   -a\gamma_5a^{-4}\delta_{x,y}.
\end{eqnarray}

One might think that a suitable modification of the chiral 
projectors
would remedy this CP breaking. But this is impossible with the 
standard CP transformation law~(\ref{cptrf}). The 
CP invariant chiral fermion action 
with a local and 
doubler-free Ginsparg-Wilson operator inevitably 
contains singularities and hence 
is non-local. We will show this below~\cite{ffis}. 
This proof will be performed in two steps.     
At the first step, we show that the projection operators 
are limited to particular forms if we require the CP symmetry 
in the chiral action under rather mild assumptions for 
the projection operators. At the second step, we show that 
these forms of the projection operators for 
local and doubler-free Ginsparg-Wilson operators are singular. 

The projection operators used in 
this section are defined as follows.
Denote the operators $\tilde{U}$ and $V$, which are regular in 
$\gamma_5$ and $H$ and satisfy 
\begin{equation}
\widetilde UH+HV=0,
\label{eq:one}
\end{equation}   
and 
\begin{eqnarray}
\label{uvsquare}
&&V^2=1,\qquad \left(\gamma_5\tilde{U}\gamma_5\right)^2=1,  
\end{eqnarray}     
we may then define the projection operators as   
\begin{eqnarray}
&&P_\mp =\frac{1}{2}(1\mp V),\qquad \bar{P}_\pm=
\frac{1}{2}(1\pm \gamma_5\tilde{U}\gamma_5).
\end{eqnarray} 
These projection operators decompose the space on which the Dirac 
operator acts as follows,
\begin{eqnarray}
&&D=\bar{P}_+ D P_- + \bar{P}_- D P_+.
\end{eqnarray}

Now we require the proper transformation property under CP, which 
is ensured by 
\begin{eqnarray}
\label{propercp}
&& WP_\mp(U^{CP})W^{-1}=\bar{P}_\pm(U)^T,\qquad
W\bar{P}_\pm(U^{CP})W^{-1}=P_\mp(U)^T.
\end{eqnarray} 
In terms of $\tilde{U}$, $V$ and $B=\gamma_5 W =\gamma_5\gamma_2$, 
these transformation 
properties are written as follows
\begin{eqnarray}
&&BV(U^{CP})B^{-1}=-\tilde{U}(U)^T,\qquad
B\tilde{U}(U^{CP})B^{-1}=-V(U)^T.
\end{eqnarray}
In fact, the chiral fermion action with the projection operators 
satisfying~(\ref{propercp}) is invariant under CP.

Then one can confirm that the projection operators with 
the above CP transformation properties~(\ref{propercp}) is limited 
to the following forms:
\begin{eqnarray}
\label{plpbarl}
P_\mp=\frac{1}{2}(1\mp \Gamma_{5}/\Gamma),\qquad 
\bar{P}_\pm=\frac{1}{2}(1\pm \gamma_{5}\Gamma_{5}\gamma_{5}
/\Gamma),
\end{eqnarray}
where 
\begin{equation}
\Gamma=\sqrt{\Gamma^{2}_{5}}=\sqrt{(\gamma_{5}\Gamma_{5}
\gamma_{5})^{2}}=\sqrt{1-H^{2}f^{2}(H^{2})}.
\end{equation} 
Let us prove this by using the following statement.\\
\\
{\bf Statement}:
If the operators $\widetilde U$ and $V$ are regular in 
$\gamma_5$ and $H$ and satisfy
\begin{equation}
\widetilde UH+HV=0,
\label{eq:one}
\end{equation}
and
\begin{equation}
B\widetilde UB^{-1}=-V^T,\quad BVB^{-1}=-\widetilde U^T,
\label{eq:two}
\end{equation}
for a {\it generic} $H$, then they are of the form
\begin{equation}
\widetilde U=V=\Gamma_5h(H^2),
\label{eq:three}
\end{equation}
with a regular function $h(H^2)$.
\\
{\bf Proof}:
Since $\gamma_5=\Gamma_5+Hf(H^2)$, $\widetilde U$ and $V$ are 
regular also in $\Gamma_5$ and $H$. Noting 
$\Gamma_5^2=1-H^2f^2(H^2)$, the most general form
of $\widetilde U$ reads
\begin{equation}
\widetilde U=g(H)+\Gamma_5h(H^2)+\Gamma_5Hk(H^2).
\label{eq:seventeen}
\end{equation}
This implies
\begin{equation}
B\widetilde UB^{-1}=-[g(H)+\Gamma_5h(H^2)-\Gamma_5Hk(H^2)]^T,
\label{eq:five}
\end{equation}
from $BHB^{-1}=-H^T$, $B\Gamma_5B^{-1}=-\Gamma_5^T$ and 
$\Gamma_5H+H\Gamma_5=0$.
eqs.~(\ref{eq:two}) and (\ref{eq:five}) imply
\begin{equation}
V=g(H)+\Gamma_5h(H^2)-\Gamma_5Hk(H^2),
\end{equation}
and thus eq.~(\ref{eq:one}) imposes
\begin{equation}
Hg(H)+\Gamma_5H^2k(H^2)=0.
\end{equation}
The matrix element of this equation between 
$\widetilde\varphi_n(x)$
and $\varphi_n(x)$ reads
\begin{equation}
(\widetilde\varphi_n,[Hg(H)+\Gamma_5H^2k(H^2)]\varphi_n)
=\sqrt{1-\lambda_n^2f^2(\lambda_n^2)}\lambda_n^2k(\lambda_n^2)=0.
\end{equation}
This shows that $k(x)$ must have zero at $x=\lambda_n^2$, 
but this is impossible for a generic $H$ unless $k(x)=0$. 
Similarly, we 
have $g(H)=0$ and obtain eq.~(\ref{eq:three}).
\\

On the basis of this statement, one can construct the 
modified chiral operators
\begin{equation}
\Gamma_{5}/\Gamma, \ \ \ \gamma_{5}\Gamma_{5}\gamma_{5}/\Gamma,
\end{equation} 
recalling~(\ref{uvsquare}) and we obtain~(\ref{plpbarl}) as 
projection operators. 
In this construction, we assumed that $h(H^2)$ exhibits the 
most favorable property, namely, has no zeroes. 
Now the projection operators are rather symmetric and it seems 
that we have succeeded in constructing CP symmetric formulation.
However, these projection operators with local and doubler-free 
Dirac operators are singular. We will next show this.
For this proof, we prove the following theorem:\\
\\
{\bf Theorem}:
For any lattice operator $D$ defined by the algebraic 
relation~(\ref{generalGW}), which 
is local (i.e., analytic in the entire Brillouin zone)
and free of species doubling, the operator 
$1/(\gamma_{5}\Gamma_{5})$ is singular inside the Brillouin 
zone and $\Gamma^{2}_{5}=1-H^{2}f^{2}(H^{2})$ has at least one 
zero inside the Brillouin zone. 
\\
{\bf Proof}:
We consider the action:
\begin{equation}
S_F=a^4\sum_{x}\bar{\psi}(x)D
\psi(x),
\end{equation}
which is invariant under the lattice chiral transformation
\begin{equation}
\delta\psi=i\epsilon\hat{\gamma}_{5}\psi, \ \ 
\delta\bar{\psi}=\bar{\psi}i\epsilon\gamma_{5}, 
\end{equation}
where $\hat{\gamma}_5=\gamma_5-2Hf(H^2)$. 
If one considers the field re-definition
\begin{equation}
\psi^{\prime}=\gamma_{5}\Gamma_{5}\psi,\ \ \ 
\bar{\psi}^{\prime}=\bar{\psi},
\end{equation}
the above action is written as 
\begin{equation}
\label{nonlocalaction}
S_F=a^4\sum_{x}
\bar{\psi}^{\prime}(x)D\frac{1}{\gamma_{5}\Gamma_{5}}
\psi^{\prime}(x),
\end{equation}
which is invariant under the naive chiral transformation:
\begin{eqnarray}
\delta\psi^{\prime}&=&\gamma_{5}\Gamma_{5}\delta\psi
=\gamma_{5}\Gamma_{5}i\epsilon\hat{\gamma}_{5}\psi
=i\epsilon\gamma_{5}\psi^{\prime},\nonumber\\
\delta\bar{\psi}^{\prime}&=&\bar{\psi}^{\prime}i\epsilon
\gamma_{5}, 
\end{eqnarray}
where we used the following relation
\begin{eqnarray}
\label{gGhatg}
\gamma_{5}\Gamma_{5}\hat{\gamma}_{5}&=&\gamma_{5}2\Gamma^{2}_{5}
-\gamma_{5}\Gamma_{5}\gamma_{5}=\gamma_{5}(\gamma_{5}\Gamma_{5}
+\Gamma_{5}\gamma_{5})-\gamma_{5}\Gamma_{5}\gamma_{5}\nonumber\\
&=&\gamma_{5}(\gamma_{5}\Gamma_{5}).
\end{eqnarray}

This chiral symmetry implies the relation
\begin{equation}
\label{gDgGsym}
\{\gamma_{5},D\frac{1}{\gamma_{5}\Gamma_{5}} \}=0.
\end{equation}
We here recall the conventional no-go theorem in the form of 
Nielsen and Ninomiya~\cite{NNtheorem}, 
which states in view of (\ref{nonlocalaction})
and (\ref{gDgGsym}) that\\
(i) If the operator $D$ is local and if
$1/(\gamma_{5}\Gamma_{5})$ is analytic in the entire 
Brillouin zone, the operator $D$ contains the species 
doubling. The simplest choice $f(H^{2})=0$ and thus 
$\Gamma_{5}=\gamma_{5}$ is included in this  case.\\
(ii)If the operator $D$ is local and free of species doubling,
then the operator $\gamma_{5}\Gamma_{5}$ is also local by 
its construction. But the operator 
$1/(\gamma_{5}\Gamma_{5})$ cannot be analytic in 
the entire Brillouin zone, which in turn suggests that 
\begin{equation}
\Gamma^{2}_{5}=1-H^{2}f^{2}(H^{2})=0,
\end{equation}
has solutions inside the Brillouin zone. These properties are 
actually proved for vanishing gauge field.\\
\\
Therefore, the projection operators~(\ref{plpbarl}) 
 inevitably contain  
singularities in the modified 
chiral operators $\Gamma_{5}/\Gamma$ and
 $\gamma_{5}\Gamma_{5}\gamma_{5}/\Gamma$.
\footnote{As a explicit example of 
this singularities, for $H$ satisfying the general Ginsparg-Wilson
relation with $f(H^2)=H^{2k}$, we have
\begin{eqnarray}
&&\frac{\Gamma_5}{\Gamma}\simeq \frac{\gamma_5\left(\gamma_\mu 
\sin ap_\mu/a\right)}{\sqrt{\sum_\mu\sin^2 ap_\mu/a^2}},
\end{eqnarray}
for $H^2\simeq 1$~\cite{Fujikawa:2001ka}. The singularities appear 
just on top of the would-be species doublers in the case of 
free-fermions and also for the topological modes 
in the presence of 
instantons.} These projection 
operators~(\ref{plpbarl}) also become 
singular in the presence of topologically non-trivial 
gauge fields, since the massive modes $N_{\pm}$ in~(\ref{GPsi})
inevitably appear as is indicated  by the chirality sum 
rule~(\ref{chiralitysumrule}).

In summary, the chiral fermion action with the local and 
doubler-free Ginsparg-Wilson operator, which is invariant under 
the standard CP transformation(\ref{cptrf}), 
contains the singularities and 
thus is non-local. Therefore it is impossible to maintain the 
manifest CP invariance of the action in L\"{u}scher's 
formulation.\footnote{This means 
however that we can keep the manifest 
CP invariance, if we ignore the singularities associated 
with $\Gamma_5=0$. But, note that the exclusion of the modes 
$\Gamma_5\varphi_n(x)=0$ in all the topological sectors is 
a non-local operation.}      
This generalizes the analysis of 
Hasenfratz~\cite{Hasenfratz:2001bz} in 
a more abstract setting.
The above CP 
breaking is thus regarded as an inherent feature of 
this formulation.
However, it should be noted that our analysis does not
show how serious the complications associated with CP
symmetry is in the actual applications 
of lattice regularization. For the breaking of CP symmetry, 
it could be similar to the breaking of Lorentz symmetry for
finite lattice spacing $a$; it may well be restored in a 
suitable continuum limit. Nevertheless it must keep 
in mind that exact and manifest CP is not implemented for a 
general Ginsparg-Wilson operator. 
Since the chirality constraint~(\ref{eq:twoxeleven}) is 
very fundamental
and it influences the construction of the fermion 
integration measure,
one might then worry that the CP breaking emerges in many 
other places
and an analysis of CP violation (in the conventional sense) 
with this
formulation would be greatly interfered. 
It is thus very important to precisely
identify where the effects of the above CP breaking inherent 
in this
formulation appear.

In the following sections we calculate the fermion generating 
functional and examine the breaking of CP symmetry~\cite{fiss}. 
Our strategy to analyze the CP breaking is as follows: 
We first determine the general structure of the fermion 
generating functional by
using a convenient auxiliary basis. Then using an 
argument of the change
of basis and a property of the measure term, we find the CP
transformation law of the generating functional.
\footnote{For an analysis of CP breakings in the fermion generating 
functional, we use the  
$t-$dependent projection operators~(\ref{eq:twoxnine})
in following sections. }

\section{Fermion generating functional}\label{fergen}

\subsection{Generating functional with an auxiliary basis}
We assume that a basis $\{v,\overline v\}$ is an ideal basis, 
i.e. that a measure current derived from the basis 
$\{v,\overline v\}$ satisfies three conditions in L\"{u}scher's 
reconstruction theorem as discussed in the previous chapter.   
To analyze the fermion generating functional for 
$\{v,\overline v\}$, $Z_{\rm F}^{\{v,\overline v\}}$, 
we introduce an auxiliary basis $\{w,\overline w\}$.  
For these two choices 
of basis, $\{v,\overline v\}$ and~$\{w,\overline w\}$,
we have
\begin{equation}
   Z_{\rm F}^{\{v,\overline v\}}[U,\eta,\overline\eta;t]
   =e^{i\theta[U;t]}
   Z_{\rm F}^{\{w,\overline w\}}[U,\eta,\overline\eta;t],
\label{eq:twoxnineteen}
\end{equation}
where the phase is given by the Jacobian factor 
for the change of basis
\begin{eqnarray}
   e^{i\theta[U;t]}&=&\det{\cal Q}\det\overline{\cal Q}
\nonumber\\
   &=&\exp(\Tr\ln{\cal Q}+\Tr\ln\overline{\cal Q}),
\label{eq:twoxtwenty}
\end{eqnarray}
where ${\cal Q}_{jk}=(w_j,v_k)$ and~$\overline{\cal Q}_{jk}=%
(\overline v_j^\dagger,\overline w_k^\dagger)$. Note that the phase
depends only on the link variable. 
Under an infinitesimal variation of the 
link variable,
the variation of the phase is given by\footnote{This is derived from
\begin{eqnarray}
   i\theta[U;t]+i\delta_\xi\theta[U;t]
   &=&\Tr\ln(w_j+\delta_\xi w_j,v_k+\delta_\xi v_k)
   +\Tr\ln(\overline v_j^\dagger+\delta_\xi\overline v_j^\dagger,
   \overline w_k^\dagger+\delta_\xi\overline w_k^\dagger)
\nonumber\\
   &=&i\theta[U;t]-i{\mathfrak L}_\xi^{\{v,\overline v\}}[U;t]
   +i{\mathfrak L}_\xi^{\{w,\overline w\}}[U;t].
\label{eq:twoxtwentytwo}
\end{eqnarray}
}
\begin{equation}
   \delta_\xi\theta[U;t]
   =-{\mathfrak L}_\xi^{\{v,\overline v\}}[U;t]
   +{\mathfrak L}_\xi^{\{w,\overline w\}}[U;t],
\label{eq:twoxtwentythree}
\end{equation}
where the ``measure term'' is defined by
\begin{equation}
   {\mathfrak L}_\xi^{\{v,\overline v\}}[U;t]
   =i\sum_j(v_j,\delta_\xi v_j)
   +i\sum_k(\delta_\xi\overline v_k^\dagger,\overline v_k^\dagger).
\label{eq:twoxtwentyfour}
\end{equation}

The relation~(\ref{eq:twoxnineteen}) shows that we may use any
basis~$\{w,\overline w\}$ as an intermediate tool in
analyzing~$Z_{\rm F}^{\{v,\overline v\}}$, if 
$e^{i\theta[U;t]}$
and~$Z_{\rm F}^{\{w,\overline w\}}$ are properly treated. 
A particularly
convenient basis is provided by the eigenfunctions of 
the hermitian
operator $D^\dagger D=H^2/a^2$:
\begin{equation}
   D^\dagger Du_j(x)
   ={1\over a^2}H^2u_j(x)={\lambda_j^2\over a^2}u_j(x),
\qquad
   \lambda_j\geq0,
\label{eq:threexone}
\end{equation}
and their appropriate projection
\begin{equation}
   w_j(x)=P_-^{(t)}u_j(x),\qquad(w_j,w_k)=\delta_{jk}.
\label{eq:threextwo}
\end{equation}
Note that $D^\dagger D$ and $P_-^{(t)}$ commute. 
To consider this eigenvalue problem, it is better to
consider first the eigenvalue problem of $H$.
The properties of these eigenfunctions are summarized
in Appendix~\ref{eigenh}.
Then going back to our original problem (\ref{eq:threexone}), 
it is obvious
that eigenfunctions are given by the 
eigenfunctions of $H$, by
identifying $\varphi_j\leftrightarrow\varphi_n$ and
$\lambda_j\leftrightarrow|\lambda_n|$ (so $\lambda_j$ is 
doubly
degenerated). To find appropriate components for
eq.~(\ref{eq:threextwo}), we have to know the action 
of chiral
projectors on these eigenfunctions $\varphi_n$. 
From above, we have

(i)~For zero modes, we simply have
\begin{equation}
   \gamma_5^{(t)}\varphi_0^\pm=\gamma_5\varphi_0^\pm
=\pm\varphi_0^\pm,
\end{equation}
and thus
\begin{equation}
   P_\pm^{(t)}\varphi_0^\pm=\pm\varphi_0^\pm.
\end{equation}

(ii) For modes with $\lambda_n\neq0$ and $\lambda_n
\neq\pm\Lambda$,
\begin{eqnarray}
   &&\gamma_5^{(t)}\pmatrix{\varphi_n\cr\widetilde
\varphi_n\cr}
\\
   &&={1\over\sqrt{1+t(t-2)\lambda_n^2f^2(\lambda_n^2)}}
   \pmatrix{(1-t)\lambda_nf(\lambda_n^2)
             &\sqrt{1-\lambda_n^2f^2(\lambda_n^2)}\cr
             \sqrt{1-\lambda_n^2f^2(\lambda_n^2)}
             &-(1-t)\lambda_nf(\lambda_n^2)\cr}
   \pmatrix{\varphi_n\cr\widetilde\varphi_n\cr}.
\nonumber
\end{eqnarray}
Since this is a traceless matrix whose determinant 
is~$-1$, the
eigenvalues of $\gamma_5^{(t)}$ in this subspace are 
$+1$ and $-1$. This
shows that one linear combination of $\varphi_n$ and
$\widetilde\varphi_n$ is annihilated by $P_+^{(t)}$ 
and the orthogonal
combination is annihilated by $P_-^{(t)}$. 
Therefore we can take
suitable linear combinations of $\varphi_n$ 
and~$\widetilde\varphi_n$
such that
\begin{equation}
   P_\pm^{(t)}\varphi_n^\pm(x)=\varphi_n^\pm(x),\qquad
   (\varphi_n^\pm,\varphi_m^\pm)=\delta_{nm}.
\end{equation}

(iii) For the modes with $\lambda_n=\pm\Lambda$, we have
\begin{equation}
   \gamma_5^{(t)}\Psi_\pm(x)=
\cases{\mp\Psi_\pm(x),&for $t>1$,\cr
   \pm\Psi_\pm(x),&for $t<1$.\cr}
\end{equation}

From the above analysis, we see that the
following vectors have an appropriate chirality as~$w_j$:
\begin{eqnarray}
   &&\varphi_0^-(x),\qquad\lambda_j=0,
\nonumber\\
   &&\varphi_j^-(x),\qquad\lambda_j>0,\qquad\lambda_j\neq\Lambda,
\nonumber\\
   &&\cases{\Psi_+(x),& for $t>1$,\cr
            \Psi_-(x),& for $t<1$,\cr}\qquad\lambda_j=\Lambda,
\label{eq:threexthree}
\end{eqnarray}
where the number~$\Lambda$ is a solution of $\Lambda f(\Lambda^2)=1$.

As for the vectors $\overline w_k$, we may adopt the left 
eigenfunctions
of the hermitian operator~$DD^\dagger$.\footnote{Incidentally, the
Ginsparg-Wilson relation implies that
$DD^\dagger=D^\dagger D$.} For non-zero
eigenvalues, as is well-known, there is a one-to-one correspondence
between the eigenfunctions of $D^\dagger D$ and $DD^\dagger$:
\begin{equation}
   \overline w_j(x)={a\over\lambda_j}w_j^\dagger D^\dagger(x).
\label{eq:threexfour}
\end{equation}
This has the proper chirality as $\overline w_k$,
$\overline w_k\overline P_+^{(t)}=\overline w_k$. The zero-modes
of~$DD^\dagger$ cannot be expressed in this way and we may use instead
\begin{equation}
   \varphi_0^{+\dagger}(x).
\label{eq:threexfive}
\end{equation}
Then eqs.~(\ref{eq:threexfour}) and~(\ref{eq:threexfive}) span a
complete set in the constrained
space\\
$\overline w_k\overline P_+^{(t)}=\overline w_k$.

Once having specified basis vectors $\{w,\overline w\}$, it is
straightforward to perform the integration
in~$Z_{\rm F}^{\{w,\overline w\}}$. After some calculations, we have
\begin{eqnarray}
   &&Z_{\rm F}^{\{w,\overline w\}}[U,\eta,\overline\eta;t]
   =\left({\Lambda\over a}\right)^N
   \prod^{n_-}\left[a^4\sum_x\overline\eta(x)\varphi_0^-(x)\right]
   \prod^{n_+}\left[a^4\sum_x\varphi_0^{+\dagger}(x)\eta(x)\right]
   \nonumber\\
&&\qquad\qquad\qquad\qquad
   \times\prod_{\lambda_j>0\atop\lambda_j\neq\Lambda}
   \left({\lambda_j\over a}\right)\,
   \exp\left[a^8\sum_{x,y}\overline\eta(x)G^{(t)}(x,y)
   \eta(y)\right],
\label{eq:threexsix}
\end{eqnarray}
up to the over-all sign $\pm1$ which depends on the ordering in the 
measure $\prod_j\rmd c_j\prod_k\rmd\overline c_k$.\footnote{This sign
factor can be absorbed into the phase~$\theta[U;t]$ without loss of
generality.} In this expression, the Green's function has been defined
by
\begin{equation}
   DG^{(t)}(x,y)=\overline P_+^{(t)}(x,y)
   -\sum^{n_+}\varphi_0^+(x)\varphi_0^{+\dagger}(y),
\label{eq:threexseven}
\end{equation}
or more explicitly,
\begin{equation}
   G^{(t)}(x,y)=\sum_{\lambda_j>0}{a^2\over\lambda_j^2}\,
   \varphi_j^-(x)\varphi_j^{-\dagger}D^\dagger(y).
\label{eq:threexeight}
\end{equation}
The number of zero-modes $\varphi_0^-$ ($\varphi_0^+$) has been denoted
by $n_-$ ($n_+$) and
\begin{equation}
   N=\cases{N_+,&for $t>1$,\cr
            N_-,&for $t<1$,\cr}
\label{eq:threexnine}
\end{equation}
where $N_+$ and $N_-$ stand for the numbers of eigenfunctions $\Psi_+$
and~$\Psi_-$, respectively (see Appendix~B). Since eigenvalues
$\lambda_j$ are gauge invariant and eigenfunctions $\varphi_j$ can be
chosen to be gauge covariant, the above $Z_{\rm F}^{\{w,\overline w\}}$
is manifestly gauge invariant for gauge covariant external sources.
However, this~$Z_{\rm F}^{\{w,\overline w\}}$ as it stands cannot be
interpreted as the generating functional for the Weyl 
fermion, as we
will explain shortly (rather it is regarded as 
representing a half of
the Dirac fermion).

\subsection{$Z_{\rm F}^{\{v,\overline v\}}$}

{}From eqs.~(\ref{eq:twoxnineteen}) and~(\ref{eq:threexsix}), the
general structure of the fermion generating functional is given by
(omitting the superscript ${\{v,\overline v\}}$)
\begin{eqnarray}
   &&Z_{\rm F}[U,\eta,\overline\eta;t]
   =e^{i\theta[U;t]}\left({\Lambda\over a}\right)^N
   \prod^{n_-}\left[a^4\sum_x\overline\eta(x)\varphi_0^-(x)\right]
   \prod^{n_+}\left[a^4\sum_x\varphi_0^{+\dagger}(x)\eta(x)\right]
   \nonumber\\
&&\qquad\qquad\qquad\qquad
   \times\prod_{\lambda_j>0\atop\lambda_j\neq\Lambda}
   \left({\lambda_j\over a}\right)\,
   \exp\left[a^8\sum_{x,y}\overline\eta(x)G^{(t)}(x,y)
   \eta(y)\right],
\label{eq:threexsixteen}
\end{eqnarray}
where the variation of the phase $\theta[U;t]$ is given by
eq.~(\ref{eq:twoxtwentythree}) and 
the measure term for the auxiliary
basis will be calculated later. We see that the vital
characterization as the {\it chiral\/} theory is contained 
in the
phase~$\theta[U;t]$ which may be computed, only after 
finding the ideal
basis ${\{v,\overline v\}}$ (or the associated measure). 
For our
discussion of CP breaking, however, only a certain 
property of the
measure term ${\mathfrak L}_\xi^{\{v,\overline v\}}[U;t]$ 
will turn out
to be sufficient.

\section{CP transformed generating functional}

\subsection{Generating functional and CP transformation}

We first note
\begin{eqnarray}
   &&\|1-R[U_{ij}^{\rm CP}(x)]\|
   =\|1-R[U_{ij}(\bar x-a\hat i-a\hat j)]\|,
\nonumber\\
   &&\|1-R[U_{i4}^{\rm CP}(x)]\|=\|1-R[U_{i4}(\bar x-a\hat i)]\|,
\label{eq:fourxone}
\end{eqnarray}
according to the CP transformation law of the plaquette variables in
Appendix~\ref{conventions}. 
Thus, if $U$ is an admissible configuration, so is
$U^{\rm CP}$; CP preserves the admissibility. Note however that $U$
and~$U^{\rm CP}$ may belong to different topological sectors in general.
In fact, the index~$n_+-n_-$~is opposite
for~$U$ and for~$U^{\rm CP}$:
\begin{equation}
   \Tr\Gamma_5(U)=n_+-n_-=-\Tr\Gamma_5(U^{\rm CP}),
\label{eq:fourxtwo}
\end{equation}
from (\ref{variousCP}).

\begin{figure}[htbp]
\begin{center}
\includegraphics{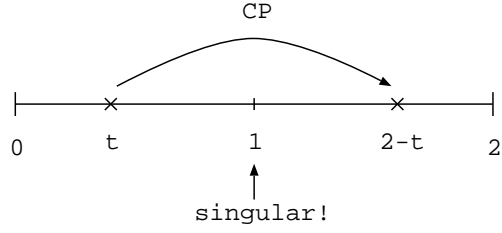}
\caption{One-parameter family of chiral gauge theories}
\label{familycgt}
\end{center}
\end{figure}

Now, let us consider the CP transformed generating functional%
\footnote{It is possible to write this formula as
\begin{eqnarray}
   &&Z_{\rm F}[U^{\rm CP},-W^{-1}\overline\eta^T,\eta^TW;t]
\label{eq:fourxfive}
\\
   &&=\int{\rm D}[\psi^{\rm CP}]{\rm D}[\overline\psi^{\rm CP}]\,
\nonumber\\   
&&\qquad \exp\biggl\{-a^4\sum_x[
   \overline\psi^{\rm CP}(x)D(U^{\rm CP})\psi^{\rm CP}(x)
   +\overline\psi^{\rm CP}(x)W^{-1}\overline\eta^T(x)
   -\eta^T(x)W\psi^{\rm CP}(x)]\biggr\}.
\nonumber
\end{eqnarray}
One thus sees that there are two possible sources of CP violation: An
explicit breaking in the action and an anomalous breaking in the path
integral measure.}
\begin{equation}
   Z_{\rm F}[U^{\rm CP},-W^{-1}\overline\eta^T,\eta^TW;t]
   =\int{\rm D}[\psi]{\rm D}[\overline\psi]\,e^{-S_{\rm F}},
\label{eq:fourxsix}
\end{equation}
where
\begin{equation}
   S_{\rm F}=a^4\sum_x[\overline\psi(x)D(U^{\rm CP})\psi(x)
   +\overline\psi(x)W^{-1}\overline\eta^T(x)
   -\eta^T(x)W\psi(x)],
\label{eq:fourxseven}
\end{equation}
and~${\rm D}[\psi]{\rm D}[\overline\psi]=%
\prod_j\rmd c_j\prod_k\rmd\overline c_k$. Here the ideal basis vectors
in $\psi(x)=\sum_jv_j(x)c_j$ and~$\overline\psi(x)=%
\sum_k\overline c_k\overline v_k(x)$ are defined through the
constraints:
\begin{equation}
   P_-^{(t)}(U^{\rm CP})v_j(x)=v_j(x),
   \qquad\overline v_k(x)\overline P_+^{(t)}(U^{\rm CP})
   =\overline v_k(x).
\label{eq:fourxeight}
\end{equation}
The generating functional~(\ref{eq:fourxsix}) can then be written as
\begin{eqnarray}
   &&Z_{\rm F}[U^{\rm CP},-W^{-1}\overline\eta^T,\eta^TW;t]
\label{eq:fourxnine}
\\
   &&=\int\prod_j\rmd c_j\prod_k\rmd\overline c_k\,
   \nonumber\\
&&\qquad\qquad \exp\biggl\{-a^4\sum_x[\overline\psi'(x)D(U)\psi'(x)
   -\overline\eta(x)\psi'(x)-\overline\psi'(x)\eta(x)]\biggr\},
\nonumber
\end{eqnarray}
where
\begin{eqnarray}
   &&\psi'(x)=[\overline\psi(\bar x)W^{-1}]^T
   =\sum_k\overline c_k[\overline v_k(\bar x)W^{-1}]^T,
\nonumber\\
   &&\overline\psi'(x)=[-W\psi(\bar x)]^T
   =\sum_j[-Wv_j(\bar x)]^Tc_j.
\label{eq:fourxten}
\end{eqnarray}
Since the basis vectors in eq.~(\ref{eq:fourxten}),
\begin{equation}
   v_k'=(\overline v_kW^{-1})^T,\qquad
   \overline v_j'=(-Wv_j)^T,
\label{eq:fourxeleven}
\end{equation}
satisfy the constraints
\begin{equation}
   P_-^{(2-t)}(U)v_k'=v_k',\qquad
   \overline v_j'\overline P_+^{(2-t)}(U)
   =\overline v_j',
\label{eq:fourxtwelve}
\end{equation}
a comparison of eq.~(\ref{eq:fourxnine}) with the original generating
functional~(\ref{eq:twoxfourteen}) shows
\begin{equation}
   Z_{\rm F}[U^{\rm CP},-W^{-1}\overline\eta^T,\eta^TW;t]
   =Z_{\rm F}[U,\eta,\overline\eta;2-t].
\label{eq:fourxthirteen}
\end{equation}
Thus the sole effect of the CP transformation is given by the change of
the parameter, $t\to2-t$ (Fig.\ref{familycgt}). 
Instead of repeating the calculation 
in Sec.~\ref{fergen} 
for~$U^{\rm CP}$, it is thus enough to examine the effect of~$t\to2-t$
in eq.~(\ref{eq:threexsixteen}).\footnote{The dimensionality of
fermionic spaces before and after CP transformation is however
different,
\begin{equation}
   \Tr[\overline P_+^{(2-t)}(U)+P_-^{(2-t)}(U)]
   -\Tr[\overline P_+^{(t)}(U)+P_-^{(t)}(U)]
   =2{t-1\over|t-1|}(n_+-n_-),
\label{eq:fourxfourteen}
\end{equation}
namely, the dimensionality jumps at $t=1$ in the presence of
topologically non-trivial gauge field.}
In eq.~(\ref{eq:threexsixteen}), we note that the zero-modes
$\varphi_0^-$, $\varphi_0^+$ and the eigenvalues~$\lambda_j$ are
independent of~$t$ (see the previous section). The change $t\to2-t$, however,
causes the exchange of $\Psi_+$ and~$\Psi_-$ as shown in
eq.~(\ref{eq:threexthree}) and thus $N\to\overline N$, where
\begin{equation}
   \overline N=\cases{N_-&for $t>1$,\cr
            N_+&for $t<1$.\cr}
\label{eq:fourxfifteen}
\end{equation}
Thus, we immediately have
\begin{eqnarray}
   &&Z_{\rm F}[U^{\rm CP},-W^{-1}\overline\eta^T,\eta^TW;t]
\nonumber\\
   &&=e^{i\theta[U;2-t]}\left({\Lambda\over a}\right)^{\overline N}
   \prod^{n_-}\left[a^4\sum_x\overline\eta(x)\varphi_0^-(x)\right]
   \prod^{n_+}\left[a^4\sum_x\varphi_0^{+\dagger}(x)\eta(x)\right]
   \prod_{\lambda_j>0\atop\lambda_j\neq\Lambda}
   \left({\lambda_j\over a}\right)
\nonumber\\
   &&\qquad\qquad\qquad\qquad
   \times
   \exp\left[a^8\sum_{x,y}\overline\eta(x)G^{(2-t)}(x,y)
   \eta(y)\right],
\label{eq:fourxsixteen}
\end{eqnarray}
for the generating functional.

\subsection{CP property of the measure term}
To derive CP transformation law of the fermion generating 
functional, we must investigate the relation between 
the phase factors, $e^{\theta[U;t]}$ 
and $e^{\theta[U;2-t]}$. Since their variation 
under an infinitesimal variation 
of the link variable is described with 
the measure terms~(\ref{eq:twoxtwentythree}), 
we examine the CP properties of the measure 
terms. We first calculate the measure terms for 
the auxiliary
basis~${\mathfrak L}_\xi^{\{w,\overline w\}}$ and then 
discuss their CP properties.

\subsubsection{Measure term for the auxiliary basis}

Here, let us consider the measure term~
(\ref{eq:twoxtwentyfour}) for the
auxiliary basis. Namely,
\begin{equation}
   {\mathfrak L}_\xi^{\{w,\overline w\}}[U;t]
   =i\sum_j(w_j,\delta_\xi w_j)
   +i\sum_k(\delta_\xi\overline w_k^\dagger,
\overline w_k^\dagger).
\label{eq:threexten}
\end{equation}
Noting $H^2w_j=\lambda_j^2w_j$ and thus
\begin{equation}
   \delta_\xi\lambda_j^2=(w_j,\delta_\xi H^2w_j),
\label{eq:threexeleven}
\end{equation}
we have from eq.~(\ref{eq:threexfour}),
\begin{equation}
   (\delta_\xi\overline w_j^\dagger,\overline w_j^\dagger)
   =-{1\over2\lambda_j^2}
   (w_j,[H,\delta_\xi H]w_j)+(\delta_\xi w_j, w_j).
\label{eq:threextwelve}
\end{equation}
Taking the contribution of zero-modes into account, 
we thus have
\begin{equation}
   {\mathfrak L}_\xi^{\{w,\overline w\}}[U;t]
   =i\sum^{n_-}(\varphi_0^-,\delta_\xi\varphi_0^-)
   +i\sum^{n_+}(\delta_\xi\varphi_0^+,\varphi_0^+)
   -{i\over2}
   \sum_{\lambda_j>0}(\varphi_j^-,[H^{-1},
\delta_\xi H]\varphi_j^-).
\label{eq:threexthirteen}
\end{equation}
The last term may be written in a basis independent way
\begin{eqnarray}
   -{i\over2}\Tr'[H^{-1},\delta_\xi H]P_-^{(t)}
   &=&{i\over4}\Tr'(H^{-1}\delta_\xi H\gamma_5^{(t)}
   -\delta_\xi HH^{-1}\gamma_5^{(t)})
\nonumber\\
   &=&{i\over4}\Tr'\delta_\xi H
   (\gamma_5^{(t)}+\gamma_5^{(2-t)})H^{-1},
\label{eq:threexfourteen}
\end{eqnarray}
where $\Tr'$ denotes the trace over the subspace of 
non-zero modes of
the hermitian operator~$H$.\footnote{One thus has to 
be careful whether
the operator concerned preserves this subspace when using 
the cyclic
property of the trace.} In deriving the last line, we have 
used the
relation $H^{-1}\gamma_5^{(t)}=-\gamma_5^{(2-t)}H^{-1}$ 
being valid in
this subspace. Noting $H^{-1}\gamma_5\gamma_5
\delta_\xi H=%
D^{-1}\delta_\xi D$, we have
\begin{equation}
   {\mathfrak L}_\xi^{\{w,\overline w\}}[U;t]
   =i\sum^{n_-}(\varphi_0^-,\delta_\xi\varphi_0^-)
   +i\sum^{n_+}(\delta_\xi\varphi_0^+,\varphi_0^+)
   +{i\over4}\Tr'\delta_\xi D(\gamma_5^{(t)}
+\gamma_5^{(2-t)})D^{-1}.
\label{eq:threexfifteen}
\end{equation}

This expression shows that the auxiliary basis $w_j$ 
and~$\overline w_k$
cannot be a physically sensible one (namely, we cannot take 
them as ideal basis 
$\{v,\overline v\}=\{w,\overline w\}$) because the 
measure term is
non-local, containing the propagator~$D^{-1}$. In fact, 
we see that
$i{\mathfrak L}_\xi^{\{w,\overline w\}}$ is identical 
to (a variation
of) the main part of the imaginary part of the fermion 
effective
action~(\ref{eq:twoxtwentyseven}), when there are no 
zero-modes.
Consequently, this basis when identified as
$\{v,\overline v\}=\{w,\overline w\}$ modifies the 
physical contents of
the theory, eliminating the imaginary part. This explains 
why the
generating functional with this basis is gauge invariant, 
even if the
fermion multiplet is not anomaly-free. Nevertheless, this 
basis is
convenient as an intermediate tool as one can work out 
all the
quantities.

\subsubsection{CP properties 
of ${\mathfrak L}_\xi^{\{w,\overline w\}}$ and 
${\mathfrak L}_\xi^{\{v,\overline v\}}$}
From (\ref{eq:threexfifteen}), 
we can see easily that the measure terms for 
the auxiliary
basis~${\mathfrak L}_\xi^{\{w,\overline w\}}$ is invariant
under~$t\to2-t$:
\begin{equation}
   {\mathfrak L}_\xi^{\{w,\overline w\}}[U;2-t]
   ={\mathfrak L}_\xi^{\{w,\overline w\}}[U;t].
\label{eq:fourxseventeen}
\end{equation}

Moreover, in the previous chapter, we have observed that the 
requirements for the
ideal measure term are given by eqs.~(\ref{anomalous})
and~(\ref{integral}).\footnote{Other important 
requirements have been 
that the measure term must be local, to be consistent 
with the locality
of the theory and that it must be smooth function of 
the link variable.} The remarkable fact is that these conditions are
invariant under $t\to2-t$. Thus, if we have an ideal measure
term~${\mathfrak L}_\xi[U;t]$ which works for $U$ with respect
to~$\gamma_5^{(t)}$, then we may use the {\it same\/} 
measure term
for~$U$ with respect to~$\gamma_5^{(2-t)}$. Therefore, 
we may set
without loss of generality
\begin{equation}
   {\mathfrak L}_\xi^{\{v,\overline v\}}[U;2-t]
={\mathfrak L}_\xi^{\{v,\overline v\}}[U;t].
\label{eq:fivexten}
\end{equation}

This equality can be interpreted in a more physically 
transparent
language; this is equivalent to the CP invariance of 
the measure term.
To see this, let us recall that basis vectors 
for~$U^{\rm CP}$ with
respect to~$\gamma_5^{(t)}$ [which is specified by
eq.~(\ref{eq:fourxeight})] and basis vectors for~$U$ 
with respect
to~$\gamma_5^{(2-t)}$ [which is specified by 
eq.~(\ref{eq:fourxtwelve})]
can be related as eq.~(\ref{eq:fourxeleven}). This 
particular choice of
bases, which may always be made, leads to
\begin{equation}
   {\mathfrak L}_\xi[U^{\rm CP};t]={\mathfrak L}_\xi[U;2-t],
\label{eq:fivexeleven}
\end{equation}
and thus eq.~(\ref{eq:fivexten}) implies
\begin{equation}
   {\mathfrak L}_\xi[U^{\rm CP};t]={\mathfrak L}_\xi[U;t].
\label{eq:fivextwelve}
\end{equation}
In fact, it is physically natural to take basis vectors 
such that the
relation~(\ref{eq:fivextwelve}) holds. In the continuum 
theory, the
action of the Weyl fermion is CP invariant and the 
imaginary part of the
effective action is too (it is independent of the 
regularization chosen
and is given by the so-called
$\eta$-invariant~\cite{Alvarez-Gaume:1985di}). This 
property is shared
with our lattice transcription~(\ref{eq:twoxtwentyseven}), 
as one can
verify from CP transformation law of various 
operators.\footnote{One
should however be careful about the meaning of the
variation~$\delta_\xi$.
Under $U_\mu(x)\to U_\mu(x)+\delta_\xi U_\mu(x)$, 
the CP transformed
configuration changes as $U_\mu^{\rm CP}(x)\to 
U_\mu^{\rm CP}(x)%
+\delta_\xi U_\mu^{\rm CP}(x)$. With this 
understanding, defining
\begin{equation}
   \delta_\xi U_\mu^{\rm CP}(x)
   =a\xi_\mu^{\rm CP}(x)U_\mu^{\rm CP}(x),
\label{eq:fivexthirteen}
\end{equation}
one has
\begin{equation}
   \xi_\mu^{\rm CP}(x)=\cases{
   -U_i^{\rm CP}(x)\xi_i(\bar x-a\hat i)^*
U_i^{\rm CP}(x)^{-1},&
   for $\mu=i$,\cr
   \xi_4(\bar x)^*,&for $\mu=4$.\cr}
\label{eq:fivexfourteen}
\end{equation}
Note that $(\xi_\mu^{\rm CP})^{\rm CP}(x)=\xi_\mu(x)$ 
corresponding to
$(U_\mu^{\rm CP})^{\rm CP}(x)=U_\mu(x)$. This definition 
of the
variation implies, in particular,
$\delta_\xi D(U^{\rm CP})=W\delta_\xi D(U)^TW^{-1}$.} 
If the measure
term is not invariant under CP, it then produces 
another unphysical
source of CP breaking as eq.~(\ref{eq:twoxtwentysix}) 
shows. In other
words, the requirement~(\ref{eq:fivextwelve}) 
eliminates an unnecessary
CP violation which may result from a wrong choice of 
the fermion measure
(which might be called ``fake CP anomaly''). 
Fortunately, it is always
possible to construct the CP invariant ideal measure 
term by the average
over CP~\cite{Luscher:1999du,Luscher:2000zd}:
\begin{eqnarray}
   &&{\mathfrak L}_\xi^{\{v,\overline v\}}[U;t]
   =a^4\sum_x\xi_\mu^a(x)j_\mu^a(x)[U;t]
\nonumber\\
   &&\to{1\over2}\biggl[a^4\sum_x\xi_\mu^a(x)
j_\mu^a(x)[U;t]
   +a^4\sum_x\xi_\mu^{{\rm CP}a}(x)j_\mu^a(x)
[U^{\rm CP};t]\biggr].
\label{eq:fivexfifteen}
\end{eqnarray}
This average is possible even if $U$ and $U^{\rm CP}$ 
belong to
different topological sectors $M$ and $M^{\rm CP}$, 
because the CP
operation defines a differentiable one-to-one 
onto-mapping from $M$
to~$M^{\rm CP}$. Then CP invariance of the measure 
term is ensured and
this is equivalent to eq.~(\ref{eq:fivexten}), 
and vice versa.

From these analysis, we have
\begin{eqnarray}
\delta_\xi\theta[U^{CP};t]&=&\delta_\xi\theta[U;2-t]\nonumber\\
&=&-{\mathfrak 
L}_\xi^{\{v,\overline v\}}[U;2-t]
   +{\mathfrak L}_\xi^{\{w,\overline w\}}[U;2-t]\nonumber\\
&=&\delta_\xi\theta[U;t], 
\label{eq:fourxnineteen}
\end{eqnarray}
and, as result,
\begin{eqnarray}
\delta_\xi(\theta[U;2-t]-\theta[U;t])=0.
\end{eqnarray}
Therefore the difference, $\theta[U;2-t]-\theta[U;t]$, 
if it exists, is a
{\it constant}:
\begin{equation}
   \theta[U;2-t]=\theta[U;t]+\theta_M,
\label{eq:fourxtwenty}
\end{equation}
where the constant~$\theta_M$ is assigned for each topological
sector~$M$, $U\in M$.

\subsection{CP transformation law of the fermion generating
functional}
{}From the above analysis, we have
\begin{eqnarray}
   &&Z_{\rm F}[U^{\rm CP},-W^{-1}\overline\eta^T,\eta^TW;t]
\nonumber\\
   &&=e^{i\theta_M}\left({\Lambda\over a}\right)^{\overline N-N}
   {\exp\left[a^8\sum_{x,y}\overline\eta(x)G^{(2-t)}(x,y)
   \eta(y)\right]\over
   \exp\left[a^8\sum_{x,y}\overline\eta(x)G^{(t)}(x,y)
   \eta(y)\right]}Z_{\rm F}[U,\eta,\overline\eta;t].
\nonumber\\
\label{eq:fourxtwentyone}
\end{eqnarray}
In particular, in the vacuum sector which contains the trivial
configuration $U_{\mu 0}(x)=1$, $U_0^{\rm CP}=U_0$ and thus one has
$\theta_M=0$ (recall that the phase depends only on 
the link variable).

{}From eq.~(\ref{eq:fourxtwentyone}), we see that the CP breaking in
this formulation appears in three places: (I)~Difference in the overall
constant phase~$\theta_M$. (II)~Difference in the overall
coefficient~$(\Lambda/a)^{\overline N-N}$ (III)~Difference in the
propagator, $G^{(t)}$ and $G^{(2-t)}$. We discuss their implications in
this order:

(I)~The {\it constant\/} phase~$\theta_M$ may be absorbed into a
redefinition of the phase factor $\vartheta_M$ in
eq.~(\ref{eq:twoxtwelve}) as
\begin{equation}
   \vartheta_M\to\vartheta_M+{1\over2}\theta_M,
   \qquad\vartheta_{M^{\rm CP}}\to
   \vartheta_{M^{\rm CP}}-{1\over2}\theta_M.
\label{eq:fourxtwentytwo}
\end{equation}
Then the overall phases in $Z_{\rm F}[U]$ and $Z_{\rm F}[U^{\rm CP}]$
become identical (no ``CP anomaly'' from the path integral 
measure) and
the discussion of CP violation is reduced to how one should 
choose the
``topological phase'' $\vartheta_M$; this is a problem analogous 
to the
strong CP problem in continuum theory.

(II)~The breaking~$(\Lambda/a)^{\overline N-N}$ can also be 
absorbed
into the topological weight~${\cal N}_M$ in~
eq.~(\ref{eq:twoxtwelve}).
Namely, we may redefine
\begin{equation}
   {\cal N}_M
   \to {\cal N}_M
   \left({\Lambda\over a}\right)^{(-N+\overline N)/2},\qquad
   {\cal N}_{M^{\rm CP}}\to {\cal N}_{M^{\rm CP}}
   \left({\Lambda\over a}\right)^{(N-\overline N)/2}.
\label{eq:fourxtwentythree}
\end{equation}
This redefinition is consistent because the roles of $N$ and
$\overline N$ are exchanged under $U\leftrightarrow U^{\rm CP}$. Note
that
\begin{equation}
   -N+\overline N=\cases{n_+-n_-,&for $t>1$,\cr
                         n_--n_+,&for $t<1$,\cr}
\label{eq:fourxtwentyfour}
\end{equation}
due to the chirality sum rule~\cite{Chiu:1998bh,Fujikawa:1999ku}. (The
index~$n_+-n_-$ does not depend on~$f(H^2)$, see Appendix~B.) The
simplest CP invariant choice is then ${\cal N}_M=1$ 
for all topological
sectors. However, whether this simplest choice is consistent 
with other
physical requirements, such as the cluster decomposition, 
is another
question which we do not discuss in this thesis. Interestingly, 
this
simplest CP invariant choice is also suggested
~\cite{Suzuki:2000ku} by a
matching with the ``Majorana formulation''.

(III)~It seems impossible to remedy the breaking 
$G^{(2-t)}\neq G^{(t)}$
in the propagator. Note that the propagator is independent 
of the choice
of the basis vectors or the path integral measure. For 
the symmetric
choice~$t=1$, $\gamma_5^{(1)}=\Gamma_5/\sqrt{\Gamma_5^2}$ is plagued
with the singularity due to zero-modes of $\Gamma_5$, whose 
inevitable
presence has been 
proven under rather mild assumptions as has been discussed 
before.
However, observe that the CP breaking for $t\neq1$ is quite 
modest. For
example, when there are no zero-modes,
\begin{equation}
   G^{(2-t)}
   =P_-^{(2-t)}{1\over D}\overline P_+^{(2-t)}
   =G^{(t)}
   +{a(1-t)f(H^2)\over\sqrt{1+t(t-2)H^2f^2(H^2)}}\gamma_5,
\label{eq:fourxtwentyfive}
\end{equation}
thus the breaking term is local. In particular, for the conventional 
choice, $t=2$ and $f(H^2)=1$,
\begin{equation}
   G^{(2-t)}(x,y)
   =G^{(t)}(x,y)-a\gamma_5{1\over a^4}\delta_{x,y},
\label{eq:fourxtwentysix}
\end{equation}
and the breaking appears as an (ultra-local) contact term, as we have
noted in Subsec.~\ref{cppro}. 
It is thus expected that this breaking is safely
removed in a suitable continuum limit in the case of pure chiral gauge
theory. However, there appear additional complications when the Yukawa
coupling is included and the Higgs field acquires the expectation
value; this issue will be discussed in the following section.

In summary, the inherent CP violation in this framework emerges only in
the fermion propagator which is connected to external sources 
in the case of pure chiral gauge theory. This
implies that diagrams with external fermion lines or with a fermion
composite operator would behave differently from the naively expected
one under CP, but the vacuum polarization, for example, respects CP. As
for other possible sources of CP violation in relative topological
weight factors, the same problem appears in continuum theory also and it
is not particular to the present formulation of lattice chiral gauge
theory.

\section{CP breaking in lattice chiral gauge theories
with Yukawa couplings}\label{secyukawa}

It is straightforward to add the Yukawa coupling to the present
formulation. By introducing the
right-handed Weyl fermion and the Higgs field, we set
\begin{eqnarray}
   S_{\rm F}&=&a^4\sum_x[
   \overline\psi_L(x)D\psi_L(x)+\overline\psi_R(x)D'\psi_R(x)
\nonumber\\   
&&\qquad\quad +\overline\psi_L(x)\phi(x)\psi_R(x)
   +\overline\psi_R(x)\phi^\dagger(x)\psi_L(x)
\nonumber\\
   &&\qquad\qquad
   -\overline\psi_L(x)\eta_L(x)-\overline\eta_L(x)\psi_L(x)
   -\overline\psi_R(x)\eta_R(x)-\overline\eta_R(x)\psi_R(x)],
\nonumber\\
\label{eq:sixxone}
\end{eqnarray}
where
\begin{eqnarray}
   &&P_-^{(t)}\psi_L=\psi_L,\qquad
   \overline\psi_L\overline P_+^{(t)}=\overline\psi_L,
\nonumber\\
   &&P_+^{\prime(t)}\psi_R=\psi_R,\qquad
   \overline\psi_R\overline P_-^{\prime(t)}=\overline\psi_R.
\label{eq:sixxtwo}
\end{eqnarray}
We assume that the left-handed fermion~$\psi_L(x)$ belongs to the
representation~$R_L$ of the gauge group and the right-handed
fermion~$\psi_R(x)$ belongs to~$R_R$ (the Higgs field~$\phi(x)$
transforms as~$R_L\otimes(R_R)^*$). The gauge couplings in the Dirac
operator~$D$ ($D'$), and correspondingly in $P_-^{(t)}$
and~$\overline P_+^{(t)}$ ($P_+^{\prime(t)}$
and~$\overline P_-^{\prime(t)}$), are thus defined with respect to the
representation~$R_L$ ($R_R$).

From now, we assume a perturbative
treatment of the Higgs coupling.
Then we note the relation
\begin{eqnarray}
   &&Z_{\rm F}[U,\phi,\eta_L,\overline\eta_L,
   \eta_R,\overline\eta_R;t]
   =\int{\rm D}[\psi]{\rm D}[\overline\psi]\,e^{-S_{\rm F}}
\nonumber\\
   &&=\exp\biggl\{a^{-4}\sum_x\biggl[
   {\partial\over\partial\eta_L(x)}\overline P_+^{(t)}\phi(x)
   P_+^{\prime(t)}{\partial\over\partial\overline\eta_R(x)}
\nonumber\\   
&&
\qquad\qquad\qquad\qquad
+{\partial\over\partial\eta_R(x)}\overline P_-^{\prime(t)}
   \phi^\dagger(x)P_-^{(t)}
   {\partial\over\partial\overline\eta_L(x)}\biggr]\biggr\}
\nonumber\\
   &&\qquad\qquad\qquad\qquad\qquad\qquad\qquad\qquad\qquad\qquad
   \times Z_{\rm F}[U,0,\eta_L,\overline\eta_L,
   \eta_R,\overline\eta_R;t],\nonumber\\
\label{eq:sixxthree}
\end{eqnarray}
because the fermion integration measure refers to neither source fields
nor the Higgs field. The generating functional without the Yukawa
coupling can be analyzed as before, and we have\footnote{It is
interesting to note that topologically non-trivial (i.e.,
$n_+-n_-\neq0$, $n_+'-n_-'\neq0$) sectors also contribute to fermion
number {\it non}-violating processes in the presence of the Yukawa
coupling.}
\begin{eqnarray}
   &&Z_{\rm F}[U,0,\eta_L,\overline\eta_L,\eta_R,\overline\eta_R;t]
\nonumber\\
   &&=e^{i\theta[U;t]}\left({\Lambda\over a}\right)^N
   \prod^{n_-}\left[a^4\sum_x\overline\eta_L(x)\varphi_0^-(x)\right]
   \prod^{n_+}\left[a^4\sum_x\varphi_0^{+\dagger}(x)\eta_L(x)\right]
   \prod_{\lambda_j>0\atop\lambda_j\neq\Lambda}
   \left({\lambda_j\over a}\right)
\nonumber\\
   &&\qquad\qquad\qquad\qquad
   \times
   \exp\left[a^8\sum_{x,y}\overline\eta_L(x)G^{(t)}(x,y)\eta_L(y)
   \right]
\nonumber\\
   &&\qquad\qquad
   \times\left({\Lambda\over a}\right)^{\overline N'}
   \prod^{n_+'}
   \left[a^4\sum_x\overline\eta_R(x)\varphi_0^{\prime+}(x)\right]
   \prod^{n_-'}\left[a^4\sum_x\varphi_0^{\prime-\dagger}(x)
   \eta_R(x)\right]
\nonumber\\   
&&\qquad\qquad\qquad
   \times\prod_{\lambda_j'>0\atop\lambda_j'\neq\Lambda}
   \left({\lambda_j'\over a}\right)\,
   \exp\left[a^8\sum_{x,y}\overline\eta_R(x)G^{\prime(t)}(x,y)
   \eta_R(y)\right],
\label{eq:sixxfour}
\end{eqnarray}
as a product of left-handed and right-handed contributions. In this
expression, all quantities with the prime ($'$) are defined 
with respect
to $H'=a\gamma_5D'$ and
\begin{equation}
   D'G^{\prime(t)}(x,y)=\overline P_-^{\prime(t)}(x,y)
   -\sum^{n_-'}\varphi_0^{\prime-}(x)\varphi_0^{\prime-\dagger}(y).
\label{eq:sixxfive}
\end{equation}

By repeating the same arguments as before\footnote{Since the charge
conjugation flips the chirality as
\begin{equation}
   \overline\psi_RD'\psi_R
   =(\psi_R^TC)D_{R_R\to (R_R)^*}'(-C^{-1}\overline\psi_R^T),
\label{eq:sixxsix}
\end{equation}
and
\begin{equation}
   (\psi_R^TC)\overline P_{+,R_R\to(R_R)^*}^{(2-t)}=(\psi_R^TC),\qquad
   P_{-,R_R\to(R_R)^*}^{(2-t)}(-C^{-1}\overline\psi_R^T)
   =(-C^{-1}\overline\psi_R^T),
\label{eq:sixxseven}
\end{equation}
the right-handed fermion may be treated as the left-handed one,
belonging to the conjugate representation~$(R_R)^*$ (with the change
$t\to2-t$). In particular, the reconstruction theorem is applied with
trivial modifications.}
for~$Z_{\rm F}[U,0,\eta_L,\overline\eta_L,\eta_R,\overline\eta_R;t]$
we finally have, corresponding to eq.~(\ref{eq:fourxtwentyone}),
\begin{eqnarray}
   &&Z_{\rm F}[U^{\rm CP},\phi^\dagger,-W^{-1}\overline\eta_L^T,
   \eta_L^TW,-W^{-1}\overline\eta_R^T,\eta_R^TW;t]
\nonumber\\
   &&=e^{i\theta_M}\left({\Lambda\over a}\right)%
   ^{\overline N-N-\overline N'+N'}
   \exp\biggl\{a^{-4}\sum_x\biggl[
   {\partial\over\partial\eta_L(x)}\overline P_+^{(2-t)}\phi(x)
   P_+^{\prime(2-t)}{\partial\over\partial\overline\eta_R(x)}
\nonumber\\   
&&
\qquad\qquad\qquad\qquad\qquad\qquad
+{\partial\over\partial\eta_R(x)}\overline P_-^{\prime(2-t)}
   \phi^\dagger(x)P_-^{(2-t)}
   {\partial\over\partial\overline\eta_L(x)}\biggr]\biggr\}
\nonumber\\
   &&\qquad
   \times{\exp\biggl\{
   a^8\sum_{x,y}\biggl[\overline\eta_L(x)G^{(2-t)}(x,y)\eta_L(y)
   +\overline\eta_R(x)G^{\prime(2-t)}(x,y)\eta_R(y)\biggr]\biggr\}
   \over
   \exp\biggl\{
   a^8\sum_{x,y}\biggl[\overline\eta_L(x)G^{(t)}(x,y)\eta_L(y)
   +\overline\eta_R(x)G^{\prime(t)}(x,y)\eta_R(y)\biggr]\biggr\}}
\nonumber\\
   &&\qquad\qquad\qquad\qquad\qquad\qquad\qquad
   \times Z_{\rm F}[U,0,\eta_L,\overline\eta_L,
   \eta_R,\overline\eta_R;t].
\label{eq:sixxeight}
\end{eqnarray}
Thus we see that the effect of the CP breaking appears precisely in the
same places as before, except for the terms consisting of Yukawa
couplings connected by the propagators.\footnote{Note that the same
projection operator, $P_+^{\prime(2-t)}$ for example, appears in the
Yukawa vertex in the combination~$\phi(x)P_+^{\prime(2-t)}$ and in the
propagator of the Weyl fermion in the
combination~$P_+^{\prime(2-t)}/D'$. Consequently, it does not matter if
one says that CP is broken either by the propagator or by the Yukawa
vertex. When $R_L=R_R$, however, it is natural to combine $\psi_L$
and~$\psi_R$ into a Dirac fermion~$\psi$. In this case, the propagator
of~$\psi$ is manifestly CP invariant and the (chirally symmetric) Yukawa
vertex breaks CP.} However, when the Higgs field acquires the
expectation value, a completely new situation arises. Setting
$\phi(x)=v$, the fermion propagators read
\begin{eqnarray}
   &&{\langle\psi_L(x)\overline\psi_L(y)\rangle_{\rm F}\over
   \langle1\rangle_{\rm F}}
   =P_-^{(t)}{1\over D-v_+D^{\prime-1}v_-}\overline P_+^{(t)}(x,y),
\nonumber\\
   &&{\langle\psi_L(x)\overline\psi_R(y)\rangle_{\rm F}\over
   \langle1\rangle_{\rm F}}
   =P_-^{(t)}{1\over v_--D'v_+^{-1}D}\overline P_-^{\prime(t)}(x,y),
\nonumber\\
   &&{\langle\psi_R(x)\overline\psi_R(y)\rangle_{\rm F}\over
   \langle1\rangle_{\rm F}}
   =P_+^{\prime(t)}{1\over D'-v_-D^{-1}v_+}
   \overline P_-^{\prime(t)}(x,y),
\nonumber\\
   &&{\langle\psi_R(x)\overline\psi_L(y)\rangle_{\rm F}\over
   \langle1\rangle_{\rm F}}
   =P_+^{\prime(t)}{1\over v_+-Dv_-^{-1}D'}\overline P_+^{(t)}(x,y),
\label{eq:sixxnine}
\end{eqnarray}
where we have defined $v_+=\overline P_+^{(t)}vP_+^{\prime(t)}$
and~$v_-=\overline P_-^{\prime(t)}v^\dagger P_-^{(t)}$. One thus sees
that this time the change $t\to 2-t$ produces {\it non-local\/}
differences in the propagator. For example, in
$\langle\psi_L(x)\overline\psi_R(y)\rangle_{\rm F}/%
\langle1\rangle_{\rm F}$, the difference
$\overline P_-^{\prime(2-t)}-\overline P_-^{\prime(t)}\propto%
a(t-1)D'\gamma_5f(H^{\prime2})$ cannot cancel the
denominator~$v_--D'v_+^{-1}D$, leaving a non-local difference.%
\footnote{Although the kernel $1/(v_--D'v_+^{-1}D)(x,y)$ decays
exponentially as $|x-y|\to\infty$, this cannot be regarded as local; the
decaying rate in the lattice unit is
$\sim 1/(\sqrt{vv^\dagger}a)\to\infty$ in the continuum limit, 
because $vv^\dagger$ is kept fixed in this limit 
(i.e., $vv^\dagger$ has
the physical mass scale).} Though we expect naively 
that this breaking, even if
it is non-local, will eventually be removed in a suitable continuum
limit, a more careful study is required to confirm this
expectation.\footnote{This non-local CP breaking will persist for a
non-perturbative treatment of the Higgs coupling, though a detailed
analysis remains to be performed.} (If one forms the free Dirac-type
propagator, the CP breaking does not appear in the propagator. 
This
means that the coupling of chiral gauge fields induces 
CP breaking.)

\chapter{Conclusion and Discussion}
In this thesis we have analyzed CP breaking
in lattice chiral gauge theory, which is a result of the 
very definition
of chirality~(\ref{eq:onexone}) for the Ginsparg-Wilson
operator~\cite{Hasenfratz:2001bz}. 
First, we showed that this CP breaking is directly 
related to the basic notions of locality and absence of
species doubling in the Ginsparg-Wilson operator~\cite{ffis}.
Although the non-perturbative construction of the ideal 
path integral
measure for non-abelian chiral gauge theories has not been 
established
yet, we analyzed the CP transformation properties of 
the fermion generating functional on the basis of a working 
ansatz~\cite{fiss}. 
Our conclusion is that there
exists no ``CP anomaly'' arising from the path 
integral measure. 
The breaking of CP is thus limited to the explicit breaking 
in the action of chiral gauge theory, and it basically appears 
in the fermion propagator in the formulation with  
Ginsparg-Wilson operators. 
When the Higgs field has no vacuum expectation value or 
in pure chiral gauge theory without the Higgs field, 
it emerges as an (almost)contact term. 
In the presence of the Higgs expectation value, however,
the breaking becomes intrinsically non-local. 
We naively expect that these breakings in the propagator, 
either local or non-local, do not survive
in a suitable continuum limit, though a detailed analysis 
is needed. One might also consider the 
CP invariant action with projection operators which are singular  
on top of the would-be species doublers 
in the case of 
free fermions or on top of 
topological modes in the presence of 
instantons,  and one might hope that those singularities are not 
so serious in an appropriate continuum limit in 
some practical applications.  
But a more careful analysis is required
to make a definite conclusion about these practical issues.

Let us now take up some topics related to both of 
the CP symmetry and 
the lattice chiral symmetry, which have not been described 
in this thesis. 

\subsubsection{CP breakings of weak matrix elements}
In this thesis, we have considered CP breakings which 
appear in the fermion generating functional 
and have not investigated CP breakings in the expectation values 
of general composite operators.  
In this case, further complications could arise. 
As an example, we
comment on the computation of the kaon $B$~parameter, 
$B_K$~in {\it lattice QCD}(see, for
example, refs.~\cite{Lellouch:2000bm,Martinelli:2001yn}). 
The following
matrix element of the effective weak Hamiltonian is then 
relevant:
\begin{equation}
   \langle\overline K^0|\,\overline s\gamma_5\Gamma_5
   \gamma_\mu{1-\gamma_5\over2}\gamma_5\Gamma_5 d\,\,
   \overline s\gamma_5\Gamma_5
   \gamma_\mu{1-\gamma_5\over2}\gamma_5\Gamma_5 d\,|K^0\rangle,
\label{eq:sevenxone}
\end{equation}
where we have adopted the $O(a)$~improved
operator~\cite{Capitani:1999uz}. Since the gauge action and the
Ginsparg-Wilson action in QCD are invariant under CP,
its CP transformation:
\begin{equation}
   \langle K^0|\,\overline d\gamma_5\Gamma_5
   \gamma_\mu{1-\gamma_5\over2}\gamma_5\Gamma_5 s\,\,
   \overline d\gamma_5\Gamma_5
   \gamma_\mu{1-\gamma_5\over2}\gamma_5\Gamma_5 s\,
   |\overline K^0\rangle.
\label{eq:sevenxtwo}
\end{equation}
coincides with the lattice transcription of 
the {\it naive\/} CP
transformation of~eq.\\
(\ref{eq:sevenxone}). 
This shows that the
$O(a)$~improvement in ref.~\cite{Capitani:1999uz}, 
which eliminates
$O(a)$ chiral symmetry breakings (in the sense of 
continuum theory),
maintains the desired behavior of the amplitude 
(\ref{eq:sevenxone})
under CP.

{}From a view point of the present analysis of CP symmetry in 
the lattice chiral gauge theory 
respecting the gauge invariance, 
however, the
above $O(a)$~improved expression of the effective Hamiltonian, if
applied to off-shell amplitudes, is not completely
satisfactory.\footnote{For on-shell amplitudes such as in
eq.~(\ref{eq:sevenxone}), the amplitude is reduced to 
the one in
continuum $V-A$ theory if the equations of motion 
for quarks are used.
In this sense, eq.~(\ref{eq:sevenxone}) and other schemes 
are
consistent. We thank Martin L\"uscher for bringing 
this fact to our
attention.} 
For example, one can confirm that the right-handed component
of the $\overline s(1-\gamma_5)/2$ 
quark\footnote{The right-handed
component of an anti-quark does not couple to 
the $W$~boson either in
continuum theory or in the standard lattice 
formulation with the overlap
operator.} contributes to the above {\it weak\/} 
effective Hamiltonian
in eq.~(\ref{eq:sevenxone}), if applied to off-shell 
Green's functions.
Although this is the order $O(a)$ effect, 
this breaks $\SU(2)_L$ gauge
symmetry of electroweak interactions. 
This illustrates that great care
need to be exercised in the analysis of lattice chiral 
symmetry and CP
invariance, when 
the expectation values of general composite operators 
are considered.

\subsubsection{Domain wall fermion and modified CP transformation}
As a closely related formulation to Ginsparg-Wilson fermions, 
the domain wall fermion~\cite{Kaplan,Shamir,kikudwf}
\cite{Vranas}-\cite{kikunogu} is 
well-known. 
In one representation of the domain wall fermion in the 
infinite flavor limit, the domain wall fermion becomes 
identical to the overlap fermion~(\ref{overlap}). 
In this case, one may  
expect that the conflict 
between CP symmetry and chiral theory naturally persists.      
Recently, in~\cite{fujisuzu}, the conflict with CP symmetry 
have been shown in a formulation of the domain wall fermion 
where the light field variables $q$ and $\bar{q}$ together 
with Pauli-Villars fields $Q$ and $\bar{Q}$ are 
utilized~\cite{Vranas}-\cite{kikunogu}. A modified 
form of lattice CP transformation motivated by 
the domain wall fermion has also been discussed there.    
For the left-handed Weyl fermion defined by the Dirac operator 
satisfying the general Ginsparg-Wilson relation~(\ref{generalGW}), 
this CP transformation is defined by 
\begin{eqnarray}
&&\psi_L \to \psi_L^{CP} =-W^{-1}\lb 
\bar{\psi}_L \frac{1}{\gamma_5\Gamma_5(U)}\rb^T, 
\nonumber\\
&&\bar{\psi}_L \to \bar{\psi}_L^{CP}=\left[\gamma_5\Gamma_5(U)
\psi_L\right]^TW,
\end{eqnarray}
where $\psi_L=\hat{P}_-\psi$ and $\bar{\psi}_L=\bar{\psi}
\bar{P}_+$ at $t=2$ in (\ref{eq:twoxnine}). An interesting feature 
of this modified lattice CP transformation 
is that one can confirm that the chiral action with Ginsparg-Wilson 
operator 
is invariant under this CP transformation. 
Moreover, one can show that the Jacobian for this modified 
CP transformation gives unity and all CP violation effects 
appear in the source terms for $\psi_L$ and $\bar{\psi}_L$. 
As a result, this analysis is perfectly 
consistent with our present analysis, 
though this modified 
CP transformation is applicable to only the   
topologically trivial sector because of the singularities in  
$1/\gamma_5\Gamma_5(U)$.  

\subsubsection{Time reversal}
We have not discussed the time-reversal in this thesis, which is the 
important discrete symmetry in chiral gauge theory.  
Since the time-reversal operator is referred to as an antilinear or 
antiunitary operator in continuum Lorentzian space-time,  
we have to begin with defining the time-reversal on the lattice, 
in other words, in the discrete Euclidean space-time.   
While one also expects the CPT invariance on the 
lattice, it is important to examine these symmetries in more 
detail.\\    
\\
For further analyses of these issues, we believe that this thesis 
will provide a good starting point.

\section*{Acknowledgments}
In the first place I would like to thank Prof. Kazuo Fujikawa 
for having led me to this interesting field and 
for his continuous encouragements and numerous discussions
and for many fruitful collaborations on this subject.  

I am also grateful 
to Prof. Hiroshi Suzuki for stimulating collaborations and 
many creative discussions and for his kindness. 

It is my pleasure to thank Prof. Atsushi Yamada, Prof. Yoshio 
Kikukawa and Dr. Tatsuya Noguchi for  
collaborations and many valuable discussions. 
I would also like to thank Dr. Oliver B\"{a}r, Dr. Kei-ichi 
Nagai, Dr. Junichi Noaki, Dr. Masataka Okamoto and Mr. Yoichi Nakayama 
for useful discussions. 

My thanks also go to Mr. Masashi Hamanaka, Mr. Yasuaki Hikida, 
Dr. Tadashi Takayanagi, Mr. Tadaoki Uesugi and Mr. Taizan Watari 
for enlightening conversations.   

I would like to thank all the members of high energy physics 
theory group at the University of Tokyo for their kindness. 

Finally, I would like to thank 
Dr. Kei Morisato for her constant support and 
encouragement at all stages of this thesis.

\appendix

\chapter{Introduction to lattice gauge theory}
In this appendix we shall review the basics of
lattice gauge theory concisely. 
First, the lattice gauge fields are introduced 
and the gauge action is constructed.
We next point out the difficulties one encounters when 
putting fermions on the lattice.
Finally, a recent progress in the treatment of the fermions
is reviewed.  
The contents are restricted to
those properties which are useful to understand this thesis and 
to theoretical issues. 
An excellent introduction to lattice gauge theory
is provided by the reviews~\cite{kogut} and 
books~\cite{creutz,montvay,rothe}.

\section{Lattice gauge fields}
Lattice gauge theory is a quantum field theory formulated
on a discrete Euclidean space time. The discrete space time
acts as a non-perturbative regularization scheme and 
enables non-perturbative studies. 
Since all components of momenta are limited to the Brillouin 
zone $[-\pi/a,\pi/a]$ at the lattice spacing $a$,
there are no ultraviolet infinities.  
\begin{figure}[htbp]
\unitlength 0.1in
\begin{picture}(44.00,32.60)(1.70,-35.30)
%
\special{pn 20}%
\special{pa 1000 290}%
\special{pa 1000 3320}%
\special{fp}%
%
\special{pn 20}%
\special{pa 1600 290}%
\special{pa 1600 3290}%
\special{fp}%
%
\special{pn 20}%
\special{pa 2190 280}%
\special{pa 2190 3290}%
\special{fp}%
%
\special{pn 20}%
\special{pa 2800 280}%
\special{pa 2800 3280}%
\special{fp}%
%
\special{pn 20}%
\special{pa 3400 270}%
\special{pa 3400 3270}%
\special{fp}%
%
\special{pn 20}%
\special{pa 600 2400}%
\special{pa 3800 2400}%
\special{fp}%
%
\special{pn 20}%
\special{sh 1}%
\special{ar 3400 1200 10 10 0  6.28318530717959E+0000}%
\special{sh 1}%
\special{ar 3400 1200 10 10 0  6.28318530717959E+0000}%
\special{sh 1}%
\special{ar 3400 1200 10 10 0  6.28318530717959E+0000}%
%
\special{pn 8}%
\special{ar 4200 1200 800 290  3.1415927 4.4626785}%
\put(41.0000,-8.2000){\makebox(0,0)[lt]{site}}%
\put(41.0000,-11.0000){\makebox(0,0)[lt]{$x=an=a(n_1,n_2,n_3,n_4)$}}%
\put(45.7000,-14.4000){\makebox(0,0)[lt]{$n\in\mbox{{\bf Z}}^4$}}%
\put(40.9000,-19.5000){\makebox(0,0)[lt]{cutoff$\>\sim \f{1}{a}$}}%
\put(40.6000,-24.1000){\makebox(0,0)[lt]{$p_\mu\in \left[-\f{\pi}{a},\f{\pi}{a} \right]$}}%
%
\special{pn 8}%
\special{pa 3990 2870}%
\special{pa 3400 2710}%
\special{fp}%
\special{sh 1}%
\special{pa 3400 2710}%
\special{pa 3459 2747}%
\special{pa 3451 2724}%
\special{pa 3470 2708}%
\special{pa 3400 2710}%
\special{fp}%
\put(40.7000,-28.3000){\makebox(0,0)[lt]{link}}%
%
\special{pn 8}%
\special{ar 1000 1500 301 301  3.6607388 4.7123890}%
%
\special{pn 8}%
\special{ar 1000 1500 301 301  1.5707963 2.5899377}%
\put(5.8000,-14.5000){\makebox(0,0)[lt]{$a$}}%
%
\special{pn 8}%
\special{pa 680 3415}%
\special{pa 1150 3415}%
\special{fp}%
\special{sh 1}%
\special{pa 1150 3415}%
\special{pa 1083 3395}%
\special{pa 1097 3415}%
\special{pa 1083 3435}%
\special{pa 1150 3415}%
\special{fp}%
\put(8.4000,-35.3000){\makebox(0,0)[lt]{$\hat{\mu}$}}%
%
\special{pn 8}%
\special{pa 490 2990}%
\special{pa 490 2510}%
\special{fp}%
\special{sh 1}%
\special{pa 490 2510}%
\special{pa 470 2577}%
\special{pa 490 2563}%
\special{pa 510 2577}%
\special{pa 490 2510}%
\special{fp}%
\put(1.7000,-26.8000){\makebox(0,0)[lt]{$\hat{\nu}$}}%
%
\special{pn 20}%
\special{pa 600 600}%
\special{pa 3810 600}%
\special{fp}%
%
\special{pn 20}%
\special{pa 600 1200}%
\special{pa 3800 1200}%
\special{fp}%
%
\special{pn 20}%
\special{pa 600 1800}%
\special{pa 3810 1800}%
\special{fp}%
%
\special{pn 20}%
\special{pa 600 3000}%
\special{pa 3810 3000}%
\special{fp}%
\end{picture}%
\caption{Lattice theory}
\end{figure}
Various physical quantities are calculated on the lattice 
and then the continuum limit is taken. When the continuum 
limit is taken, the bare 
coupling $g_0$ has to be tuned with the lattice spacing $a$ so that
physical quantities may be finite in the limit
of zero lattice spacing.

On the lattice the gauge fields are represented by
link variables $U_\mu(x)$ on the link. The link
variable is an element of the gauge 
group $G$, for example, $SU(N)$ or $U(1)$. 
The relation between 
$U_\mu(x)$ and the vector 
potential $A_\mu(x)$ is  written as
\begin{eqnarray}
&&U_\mu(x)=e^{iagA_\mu(x+a\hat{\mu}/2)}.
\end{eqnarray}
Under
gauge transformations
\begin{eqnarray}
&& U_\mu(x)\to \Lambda(x)U_\mu(x)\Lambda(x+a\hat{\mu})^{-1},
\qquad \Lambda(x)\in \mbox{G}.
\end{eqnarray}
The gauge covariant variables corresponding to the 
field strength in continuum are 
plaquette variables $U_{\mu\nu}(x)$ which are 
the product of link variables
around an elementary plaquette:
\begin{eqnarray}
&&U_{\mu\nu}(x)=U_\mu(x)U_\nu(x+a\hat{\mu})
U_\mu^\dagger (x+a\hat{\nu})U_\nu^\dagger (x).
\end{eqnarray}
The gauge action, which is called the Wilson 
action, is defined in terms of these 
plaquette
variables as
\begin{eqnarray}
\label{wilsonaction}
&&S_G=\frac{1}{g_0^2}a^4\sum_{x,\mu\neq\nu}
\Tr\left(1-U_{\mu\nu}(x)\right),
\end{eqnarray}
with $g_0$ being the bare coupling. This action
agrees with the continuum theory
in the classical continuum 
limit~\footnote{The classical continuum limit means
that the limit of zero lattice spacing 
alone is considered.}.

\section{Fermion doubling problem and 
Nielsen-Ninomiya no-go theorem}
The fermion doubling is the problem that extra fermion species
appear when one naively puts the fermion
on the lattice. In the following, we discuss 
this problem.
A naive massless Dirac fermion action is written 
as follows,\footnote{On the lattice the derivative is replaced by 
the difference operator. We define the forward difference 
operators $\partial_\mu$ and the backward difference operators as
follows, 
\begin{eqnarray}
\partial_\mu\psi(x)&=&\frac{1}{2a}\left(\psi(x+a\hat{\mu})
-\psi(x)\right),\\
\partial_\mu^*\psi(x)&=&\frac{1}{2a}\left(\psi(x)-\psi(x-a\hat{\mu})
\right).
\end{eqnarray}
} 
\begin{eqnarray}
\label{naction}
S_F&=& a^4\sum_x\bar{\psi}(x)D\psi(x)\nonumber\\
&=&a^4\sum_x\bar{\psi}(x)(\partial_\mu +\partial_\mu^*)
\psi(x)\nonumber\\
&=& a^4\sum_{x,\mu}\bar{\psi}(x)\f{\gamma_\mu}{2a}
\left(\psi(x+a\hat{\mu}) - \psi(x-a\hat{\mu})\right),
\end{eqnarray}
which is invariant under the naive chiral symmetry
\begin{eqnarray}
\label{nchiral}
\psi(x)&\to&\psi^\prime(x)=\psi(x)+i\epsilon\gamma_5\psi(x),
\nonumber\\
\bar{\psi}(x)&\to&\bar{\psi}^\prime(x)=\bar{\psi}(x)
+i\epsilon\bar{\psi}(x)\gamma_5,
\end{eqnarray}
with a real infinitesimal parameter $\epsilon$ and
see Appendix \ref{conventions} for $\gamma$-matrix conventions.
Then the propagator is given by
\begin{eqnarray}
&&<\psi(x)\bar{\psi}(y)> =  \int^{\pi/a}_{-\pi/a}
\f{dp^4}{(2\pi)^4}\, e^{ip(x-y)a}
\f{a}{i\sum_\mu \gamma_\mu\sin ap_\mu}.
\end{eqnarray}
We set $\tilde{p}_\mu = \f{i}{a}\sin ap_\mu$ and 
see the poles of this propagator in momentum space.
\begin{figure}[hbtp]
\begin{center}
\includegraphics{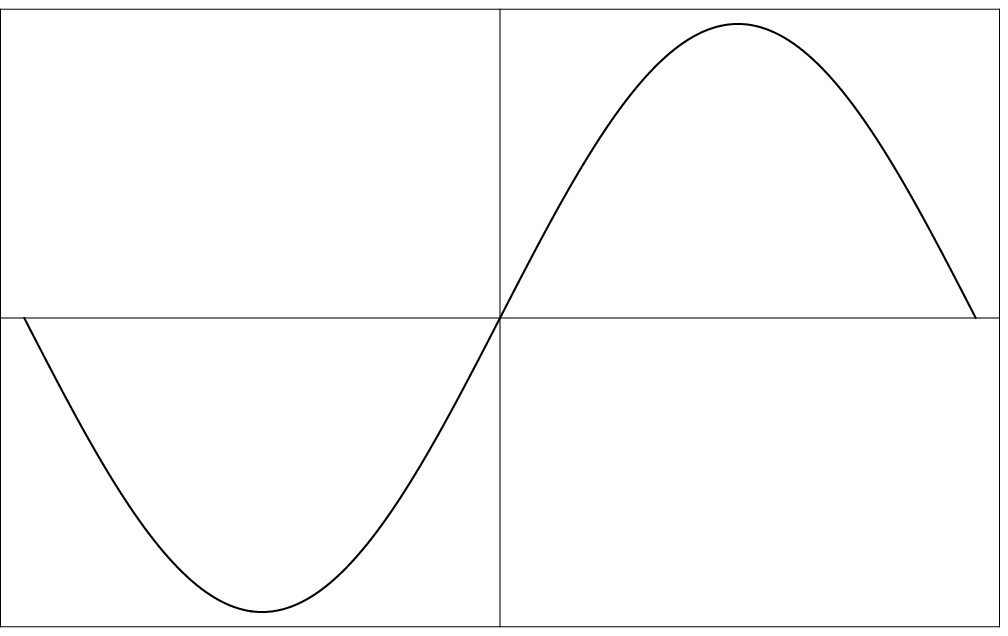}
\caption{Plot of $\tilde{p}_\mu$ versus $p_\mu$ in the
Brillouin zone}
\label{fig01}
\begin{picture}(10,10)
\put(-132,140){$-\pi/a$}
\put(122,112){$\pi/a$}
\put(158,130){$p_\mu$}
\put(-3,225){$\tilde{p}_\mu$}
\end{picture}
\end{center}
\end{figure}
By Fig.(\ref{fig01}) we easily find that 
$\tilde{p}_\mu$ has two zeros. From this,   
the propagator in 4 dimension has 16 massless poles. Hence
the above lattice theory contains sixteen species 
of fermions. The excitation around $p_\mu = (0,0,0,0)$ is 
called the physical mode and 15 modes except the physical
mode are called doubler modes. This is the well-known fermion 
doubling problem. To obtain the correct continuum 
limit, doubler modes must be eliminated.   

\subsection{Wilson fermion}\label{wilsonfermion}
In the lattice theory one has 
the freedom to add the irrelevant operator
to the action and thus there are many different actions
which have the same classical continuum limit.
The action described above has been chosen as the 
simplest one. Let us now modify 
the action~(\ref{naction}) by adding
a term with a second difference 
operator
which vanishes in the classical
continuum limit~\cite{Wilsonfermion}. Thus we have
\begin{eqnarray}
\label{Wilsonfer}
S_F^{(W)} &=& a^4 \sum_{x,\mu}\bar{\psi}(x)
(\gamma_\mu(\partial_\mu+\partial_\mu^*) 
- \f{a}{2}r
\partial_\mu\partial_\mu^*)\psi(x)\nonumber\\
&=& a^4\sum_{x,\mu}\bar{\psi}(x)\lc\f{\gamma_\mu}{2a}
(\psi(x+a\hat{\mu})
-\psi(x-a\hat{\mu}))\right.\nonumber\\
&&\left.\qquad\qquad -\f{r}{2a}(\psi(x+a\hat{\mu}) 
+ \psi(x-a\hat{\mu})
-2\psi(x))\rc, 
\end{eqnarray}
with a constant $r$ which is called the Wilson parameter.
The above action has the mass term in the momentum space
as follows,
\begin{eqnarray}
M(p)&=&\f{r}{a}\sum_\mu(1-\cos p_\mu a)\nonumber\\
&=&\lc\begin{array}{cc}
      0 & \quad\mbox{for physical mode,}\\
     C/a & \quad\mbox{for doubler modes,} 
   \end{array}
\right.
\end{eqnarray}
where $C$ is a constant which is different from each
doubler mode. Thus the doubler modes
get masses proportional to $r/a$ in virtue of the Wilson term.
As the result the doubler modes decouple in the continuum 
limit when the Wilson fermion is used.   
It is clear, however, that the Wilson
fermion action has no naive chiral 
symmetry~(\ref{nchiral})
because of the Wilson term.
This conflict between no doublers and that 
the naive chiral symmetry has been examined in detail 
and has been  summarized in the
following no-go
theorem~\cite{NNtheorem}. \\
\\
{\bf Nielsen-Ninomiya no-go theorem}\\
Lattice Dirac operator $D$ cannot have 
the following properties simultaneously.
\begin{description}
\item[(a)]  $D(x,y)$ is local.
\item[(b)] $\tilde{D}(p) = i\gamma_\mu p_\mu
+ {\cal O}(ap^2)$ for $p\ll \f{\pi}{a}$.
 (the correct behaviour in the classical continuum limit)
\item[(c)] $\tilde{D}(p)$
is invertible when $p\ne 0$ ($mod\>\f{2\pi}{a}$). 
 (no massless doublers)
\item[(d)] $\lc \g5, D\rc = 0$. (naive chiral symmetry)
\end{description}
where $\tilde{D}(p)$ is the Fourier transform of $D(x,y)$.\\
\\
The properties (a) and (b) are important for getting
the correct continuum theory in the continuum limit.
Therefore this theorem suggests that one
may remove the doubler modes at the expense of breaking
the naive chiral symmetry and it had been considered that
putting a single massless fermion on the lattice is  
difficult. However a recent progress allows a single massless 
fermion on the lattice. We discuss this in the
following section.

Now that we have defined the lattice action of
vector-like theory, let us 
construct the quantum theory by using the path integral
in the remainder of this subsection.   
We first make the Wilson fermion action~(\ref{Wilsonfer}) 
invariant under the gauge transformations
\begin{eqnarray}
&&\psi(x)\to \Lambda(x)\psi(x),\qquad \bar{\psi}(x)\to
\bar{\psi}(x)\Lambda^{-1}(x),
\end{eqnarray}
where $\Lambda(x)$ is an element of the gauge group. 
Using the link variables $U_\mu(x)$, the gauge
invariant fermion action is written as
\begin{eqnarray}
\label{gaugeWilsonfer}
S_F^{(W)}
&=&a^4\sum_{x,y,\mu}\bar{\psi}(x)\left(\sla{C}(x,y)+B(x,y)
+\frac{m_0}{a}\delta_{x,y}\right)\psi(y),\nonumber\\ 
&=&a^4\sum_{x,y,\mu}\f{1}{2a}\bar{\psi}(x)\lb
\gamma_\mu \left(U_\mu(x)\psi(x+a\hat{\mu})
-U_\mu^\dagger(x-a\hat{\mu})\psi(x-a\hat{\mu})\right)
\right.\nonumber\\
&&\left.+r\left(2\psi(x)-U_\mu(x)\psi(x+a\hat{\mu}) 
-U_\mu^\dagger(x-a\hat{\mu})\psi(x-a\hat{\mu})\right)\rb
\nonumber\\
&&+\frac{m_0}{a}\bar{\psi}(x)\psi(x),
\end{eqnarray}
where $\sla{C}=\gamma_\mu C_\mu$ and 
\begin{eqnarray}
\label{wilsondiraccb}
&&C_\mu(x,y)=\f{1}{2a}\lb \delta_{x^\mu + \hatmu a,y^\mu}
U_\mu(x)
- \delta_{x^\mu,y^\mu + \hatmu a}U_\mu^\dagger(y)\rb, \\
&&B(x,y)=\f{r}{2a}\sum_{\mu}\lb 2\delta_{x,y}-
\delta_{x^\mu + \hatmu a,y^\mu}U_\mu(x)
- \delta_{x^\mu,y^\mu + \hatmu a}U_\mu^\dagger(y)\rb,
\end{eqnarray}
with $m_0$ being the bare mass parameter.
To construct the quantum theory, we must define the 
integration measure. We adopt an invariant 
group measure, the Haar measure, $D[U]$ as the measure 
of the integration over the link variables and 
introduce Berezin integration as the 
integration over fermion fields. 
Consequently, we obtain the path integral expression
for the expectation value of ${\cal O}$:
\begin{eqnarray}
\label{vectorexp}
\langle {\cal O} \rangle&=&
\f{1}{Z}\int D[U]D[\bar{\psi}]D[\psi]{\cal O}\> e^{-S},
\nonumber\\
&&Z=\int D[U]D[\bar{\psi}]D[\psi]\> e^{-S},
\end{eqnarray}
where $S=S_G[U]+S_F^{(W)}[U,\psi,\bar{\psi}]$.
An arbitrary correlation function containing
the fermion fields and the link variables in the above formulation
 can be
calculated from the above path integral.
Of course, when we try to treat a massless fermions, 
we need a fine tuning of the parameter $m_0$ 
to keep the fermion massless in the continuum limit 
because of the breaking of chiral symmetry. 
  
\subsection{Chiral gauge theory with the 
Wilson fermion}\label{chiralWilson}
We have just discussed the vector-like theory on the lattice.
Now we consider the lattice formulation of
chiral gauge theory with the Wilson fermion.
In chiral gauge theory, the chiral symmetry is 
gauged, but the Wilson fermion action 
breaks chiral symmetry. This means that this formulation
breaks gauge invariance and thus we suffer from
the gauge degrees of freedom and must try to recover the 
gauge symmetry in the continuum limit.
We next illustrate the above situation 
by an example~\cite{Golterman:2001hr}. 
We assume that 
the fermions which have colors are left-handed.
The chiral fermion action with the Wilson fermion
is given by
\begin{eqnarray}
&&S_F=a^4\sum_{x,y,\mu}\bar{\psi}_L(x)
\sla{C}(x,y)\psi_L(y)-a
\left(\bar{\psi}_R(x)B(x,y)\psi_L(y) + h.c. \right),\nonumber\\
\end{eqnarray}
where
\begin{eqnarray}
&& \psi_L =\frac{1-\gamma_5}{2}\psi,\qquad
\bar{\psi}_L = \bar{\psi}\frac{1+\gamma_5}{2},\nonumber\\
&&\psi_R =\frac{1+\gamma_5}{2}\psi,\qquad 
\bar{\psi}_R = \bar{\psi}\frac{1-\gamma_5}{2},
\end{eqnarray}
and where $\psi_R$ is a gauge-neutral fermion introduced
only in order to
remove the doublers of the charged left-handed 
fermion $\psi_L$. After the  
gauge transformations $\psi_L(x)\to \Lambda(x)$ and 
$U_\mu(x)\to \Lambda(x)U_\mu(x)\Lambda(x+a\hat{\mu})$,
this action becomes
\begin{eqnarray}
&&S_F=a^4\sum_{x,y,\mu}\bar{\psi}_L(x)
\sla{C}(x,y)\psi_L(y)-a
\left(\bar{\psi}_R(x)\Lambda(x)B(x,y)\psi_L(y) + h.c. \right).
\nonumber\\
\end{eqnarray}  
We see that the gauge degrees of freedom $\Lambda(x)$ couple
to the fermion because the Wilson term breaks the gauge 
symmetry. The dynamics of this coupling has drastic
consequences for the fermion spectrum:for instance, the
theory becomes vector-like in the continuum 
limit~\cite{petcher}.    
It is clear that we need a more clever idea if we use
the Wilson fermion in the construction of lattice
chiral gauge theory. In this direction, two methods
have been known:one using different cutoffs for
the fermions and the link variables and another using the 
gauge 
fixing to control the gauge degrees of freedom.
In the latter method, abelian chiral gauge theory has
been constructed in the continuum limit~\cite{bochleung}.

\section{Massless fermion on the lattice}
The first breakthrough in the attempts to put  
a massless fermion on the lattice may be 
traced to the domain wall fermion~\cite{Kaplan,Shamir},
 which was followed by the overlap 
formalism~\cite{Narayanan:1993wx,Randjbar-Daemi:1995sq}
and then a remarkable identity, which first
appeared in 1982 in the paper of Ginsparg and 
Wilson~\cite{Ginsparg:1982bj},
 was re-discovered~\cite{Hasenfratz:1998ft,Neuberger:1998fp}.
This identity is the Ginsparg-Wilson relation:
\begin{eqnarray}
\label{GW}
D\g5 + \g5 D = 2a D\g5 D.
\end{eqnarray}
This relation avoids the Nielsen-Ninomiya no-go theorem
on account of the term proportional to the lattice spacing in
the right-hand side. The above relation can be
written, in terms of the kernel of Dirac operator, as follows,
\begin{eqnarray}
D^{-1}(x.y)\g5 + \g5 D^{-1}(x,y) = 2a\g5\delta_{x,y}.
\end{eqnarray}
This expression means 
that the breaking of the naive chiral symmetry
would not have the effect on the physical amplitudes.
Besides this relation represents the exact lattice 
chiral symmetry~\cite{luscherchiral}.   
The fermion action with the Ginsparg-Wilson Dirac operator 
is invariant
under the following transformation:
\begin{eqnarray}
&&\psi \rightarrow \psi^\prime = \psi + i\epsilon
\hat{\gamma}_5\psi,\nonumber\\
&&\bar{\psi}\rightarrow \bar{\psi}^\prime=\bar{\psi} 
+ i\epsilon\bar{\psi}
\g5,
\end{eqnarray}
where $\hat{\gamma}_5\equiv\g5-2a\g5 D\quad 
(\hat{\gamma}_5^2=1)$. Thus it is expected that
the formulation
with the Ginsparg-Wilson fermion has the same properties 
as the continuum theory with the chiral 
symmetry~\footnote{The other 
implications 
of the Ginsparg-Wilson relation are ~\cite{Hasenfratz:1998jp}
\begin{itemize}
 \item the formulation
with this relation does not require fine tuning
to obtain the pion massless.
\item neither the vector
nor the flavor non-singlet axial vector current need
renormalization.
\item there is no mixing between four-fermion operators
in different chiral representations.
\item the divergence of the flavor non-singlet 
axial vector current is non-anomalous, while the
flavor singlet axial vector current has the anomalous
divergence.
\end{itemize}}. 
In general, the Ginsparg-Wilson relation, however, does not
guarantee the locality of the Dirac operator and no doublers. 
Therefore it is important to find the Dirac operator
satisfying the Ginsparg-Wilson relation with 
such good properties. One of such 
Dirac operators is 
the overlap Dirac 
operator~\cite{Neuberger:1998fp}, which is 
described explicitly as~\footnote{As such a Dirac operator, 
the fixed point action~\cite{Hasenfratz:1998ft} is also
well known.}
\begin{eqnarray}
\label{overlap}
&&D = \f{1}{2a}\left(1+ D_W \f{1}{\sqrt{D^\dagger_W D_W}}
\right),
\end{eqnarray}
with $D_W$ is the Wilson-Dirac operator with the negative mass:
\begin{eqnarray}
&&D_W(x,y)=\sla{C}(x,y) 
+ B(x,y) - \f{1}{a}m_0\delta_{x,y},
\end{eqnarray}
with (\ref{wilsondiraccb}). 
The overlap Dirac operator is gauge-covariant and also 
is local and smooth function of link variables 
when the admissibility 
condition~\cite{Hernandez:1999et,Neuberger:2000pz}:
\begin{eqnarray}
\| 1-R[U_{\mu\nu}(x)]\|<\frac{1}{6(2+\sqrt{2})},
\end{eqnarray} 
is satisfied for all plaquette variables~\footnote{This
restriction on the space of link variables is 
the sufficient condition
to ensure the locality and the smoothness of the Dirac 
operator.}. The fermion propagator has the single pole
for the physical mode and is free from species doublers, when we take  
$0<m_0<2$. 
Moreover, the axial anomaly from the fermion measure
has been evaluated~\cite{kikuyama}-\cite{suzuano} in the 
continuum limit 
and also the perturbative properties
of the lattice QCD with this Dirac operator has been 
discussed~\cite{ishipertur}-\cite{Alexandrou:2000kj}. 
In addition, unitarity also has
been established on the lattice in the case of free 
fermion by showing  
that a K\"{a}ll\'{e}n-Lehmann representation of the fermion 
propagator 
has non-negative spectral density~\cite{Luscher:2000hn}. 
All analysis
clearly show that the lattice theory with the overlap 
Dirac operator reproduces the properties of massless Dirac
fermion in continuum theory well. The overlap Dirac
operator satisfies the properties of (a), (b) and (c) in
Nielsen-Ninomiya's theorem, but, of course, doesn't satisfy 
(d). 

Here, let us mention the definition of 
locality on the lattice. 
As the definition 
of locality, we have adopted 
the exponential locality, which means that 
the kernel has exponentially decaying tails at large distances:
\begin{eqnarray}
 &&\| D(x,y)\| \le C\,e^{-\gamma|x-y|/a},
\end{eqnarray}
where $\gamma$ and $C$ are constants and the norm $\| M\|$ for a 
matrix $M$ is defined by 
\begin{eqnarray}
\| M\|=\sqrt{\lambda_{max}(
M^\dagger M)},
\end{eqnarray}
where $\lambda_{max}$ is maximal eigenvalues of $M$.
The overlap Dirac operator~(\ref{overlap}) is an example of 
the Dirac operator with such a locality. 
From the view point of 
continuum limit 
this locality is as good as the strict locality.
\footnote{The strict locality means 
that the two points, $x$ and $y$, 
of the kernel of the Dirac operator $D(x,y)$ are only finite 
distances away in the lattice unit. 
The example of the Dirac operator with this 
locality is the Wilson fermion~(\ref{gaugeWilsonfer}).     
}

\subsection{Algebraic generalization of the
Ginsparg-Wilson relation}
It is important to seek alternative lattice Dirac 
operators, from a theoretical viewpoint, to
deepen the understanding of Ginsparg-Wilson fermions. 
From a numerical one, one might hope to improve 
the properties, such as chiral symmetry and locality 
of Dirac operators, compared to the overlap Dirac
operator and the fixed point action.
While this task is challenging and difficult, 
a new class of lattice Dirac operators which are based on the 
algebraically generalized Ginsparg-Wilson 
relation have been proposed~\cite{Fujikawa:2000my}. 
This general Ginsparg-Wilson relation is written as 
\begin{eqnarray}
\label{fujialgebra}
&&\gamma_5 D + D\gamma_5 =2a D(a\gamma_5 D)^{2k}\gamma_5 D,
\end{eqnarray}
where $k$ stands for a non-negative integer, 
and $k=0$ corresponds to the ordinary 
Ginsparg-Wilson relation. 
The solution $D$ of this general 
Ginsparg-Wilson relation~(\ref{fujialgebra}) is
given by
\begin{eqnarray}
\label{gwo2}
D&=&\frac{1}{a}\g5\lb\f{1}{2}\g5\left(1+ D_W^{(2k+1)} 
\f{1}{\sqrt{D^{(2k+1)\dagger}_W D_W^{(2k+1)}}}
\right)\rb^{1/(2k+1)},
\end{eqnarray}
where
\begin{eqnarray}
D_W^{(2k+1)}&=&i(\sla{C})^{2k+1} + (B)^{2k+1} - 
\left(\f{m_0}{a}\right)^{2k+1},\nonumber\\
\end{eqnarray}
with $C_\mu$ and $B$ in (\ref{wilsondiraccb}).
In this class of lattice Dirac operators, $k=0$ 
corresponds to the overlap Dirac 
operator.
When the parameter $m_0$ is chosen as $0<m_0<2r$ in these
operators,
the free propagators have a single massless pole and no doublers.
The locality properties have been confirmed strictly 
in the case of the free operator~\cite{fujilocal}.
In addition, it has been shown that 
the formulation with these operators
gives the correct chiral and Weyl anomalies~\cite{fujichiral,
fujipertur} and thus the topological properties of these
operators are identical to those of the standard overlap
Dirac operator. 

A salient feature of Dirac operators corresponding to 
larger $k$ is that the chiral symmetry(in the sense of continuum) 
breaking
term 
becomes more irrelevant in the sense of
Wilsonian renormalization group. 
For the operators~(\ref{gwo2}), 
in the near continuum configurations, 
we see the behavior as follows
\begin{eqnarray}
D&\simeq& i\sla{D}+ a^{2k+1}(\g5 i\sla{D})^{2k+2}, 
\end{eqnarray}
where $\sla{D}= \ga_\mu(\partial_\mu+igA_\mu)$.
The first terms in this expression stands for the leading term
in chiral symmetric terms, and the second term stands for the
leading term in chiral symmetry breaking terms. This shows
that one can improve the chiral symmetry for larger $k$. 
As another manifestation of this property, the spectrum of 
the operators with $k\ge 1$ is closer to that of 
the continuum operator
in the sense that the small eigenvalues of $D$ accumulate along
the imaginary axis (which is a 
result of taking a $2k+1$-th root)~\cite{C}, compared 
to the overlap Dirac operator for which the eigenvalues of
$D$ draw a perfect circle in the complex eigenvalue plane.  
However, the locality property becomes worse for larger $k$.  
To understand the properties of these operators better,
a detailed numerical study
of locality and chiral symmetry 
has to be performed for larger $k$.     
\\

We can thus formulate the lattice theory with 
Ginsparg-Wilson fermions\\
(either by the conventional one 
or its generalization) at the quantum level, using the same
fermion measure as discussed in Subsec.\ref{wilsonfermion}. 
In this formulation, we need no fine-tuning to 
keep the fermion massless, while we have needed in the formulation
with Wilson fermions. The formulation with Ginsparg-Wilson
fermions has enabled to calculate 
the physical quantities that the chiral
symmetry is important, for example, 
the $CP$-violation parameter in the Standard Model 
$\epsilon^\prime/\epsilon$ and has opened the possibility
to construct the chiral gauge theory on the lattice 
with exact gauge invariance. 
See Chapter.~\ref{latchiralgauge}.

\chapter{Eigenvalue problem of $H$}\label{eigenh}
In this Appendix we consider the following eigenvalue problem
for the operator $H$ satisfying the general Ginsparg-Wilson 
relation~(\ref{generalGW}):
\begin{equation}
   H\varphi_n(x)=\lambda_n\varphi_n(x),\qquad
   (\varphi_n,\varphi_m)=\delta_{nm}.
\label{eq:axone}
\end{equation}
We note
\begin{equation}
   H\Gamma_5\varphi_n(x)=-\Gamma_5H\varphi_n(x)
   =-\lambda_n\Gamma_5\varphi_n(x),
\end{equation}
and
\begin{equation}
   (\Gamma_5\varphi_n,\Gamma_5\varphi_m)
   =[1-\lambda_n^2f^2(\lambda_n^2)]\delta_{nm}.
\end{equation}
These relations show that eigenfunctions with $\lambda_n\neq0$ 
(when
$\lambda_n=0$, $\varphi_0(x)$ and $\Gamma_5\varphi_0(x)$ are not
necessarily linear-independent) and $\lambda_nf(\lambda_n^2)\neq\pm1$
come in pairs as $\lambda_n$ and~$-\lambda_n$.

We can thus classify eigenfunctions in eq.~(\ref{eq:axone}) as follows:

(i) $\lambda_n=0$ ($H\varphi_0(x)=0$). For this
\begin{equation}
   H{1\pm\gamma_5\over2}\varphi_0(x)=H{1\pm\Gamma_5\over2}\varphi_0(x)
   ={1\mp\Gamma_5\over2}H\varphi_0(x)=0,
\end{equation}
so we may impose the chirality on $\varphi_0(x)$ as
\begin{equation}
   \gamma_5\varphi_0^\pm(x)=\Gamma_5\varphi_0^\pm(x)
   =\pm\varphi_0^\pm(x).
\end{equation}
We denote the number of $\varphi_0^+(x)$ ($\varphi_0^-(x)$) as $n_+$
($n_-$).

(ii) $\lambda_n\neq0$ and $\lambda_nf(\lambda_n^2)\neq\pm1$. As shown
above,
\begin{equation}
   H\varphi_n(x)=\lambda_n\varphi_n(x),\qquad
   H\widetilde\varphi_n(x)=-\lambda_n\widetilde\varphi_n(x),
\end{equation}
where
\begin{equation}
   \widetilde\varphi_n(x)={1\over\sqrt{1-\lambda_n^2f^2(\lambda_n^2)}}
   \Gamma_5\varphi_n(x).
\end{equation}
We have
\begin{equation}
   \Gamma_5\varphi_n(x)=\sqrt{1-\lambda_n^2f^2(\lambda_n^2)}
   \widetilde\varphi_n(x),\qquad
   \Gamma_5\widetilde\varphi_n(x)=\sqrt{1-\lambda_n^2f^2(\lambda_n^2)}
   \varphi_n(x),
\end{equation}
and
\begin{equation}
   \gamma_5\pmatrix{\varphi_n\cr\widetilde\varphi_n\cr}
   =\pmatrix{\lambda_nf(\lambda_n^2)
             &\sqrt{1-\lambda_n^2f^2(\lambda_n^2)}\cr
             \sqrt{1-\lambda_n^2f^2(\lambda_n^2)}
             &-\lambda_nf(\lambda_n^2)\cr}
   \pmatrix{\varphi_n\cr\widetilde\varphi_n\cr}.
\end{equation}

(iii) $\lambda_nf(\lambda_n^2)=\pm1$, or
\begin{equation}
   H\Psi_\pm(x)=\pm\Lambda\Psi_\pm(x),\qquad\Lambda f(\Lambda^2)=1.
\end{equation}
We see
\begin{equation}
\label{GPsi}   
\Gamma_5\Psi_\pm(x)=0,
\end{equation}
and
\begin{equation}
   \gamma_5\Psi_\pm(x)=\pm\Lambda f(\Lambda^2)\Psi_\pm(x)
   =\pm\Psi_\pm(x).
\end{equation}
We denote the number of $\Psi_+(x)$ ($\Psi_-(x)$) as $N_+$ ($N_-$).
As a application of the above relations, we have
\begin{equation}
   \Tr\gamma_5=n_+-n_-+N_+-N_-.
\end{equation}
If we further note $\Tr\gamma_5=0$ on the lattice, we have 
the chirality
sum rule~\cite{Chiu:1998bh,Fujikawa:1999ku}
\begin{equation}
\label{chiralitysumrule}   
n_+-n_-+N_+-N_-=0.
\end{equation}

\chapter{Some computations}\label{computations}
\subsubsection{eqs. (\ref{eq:fivexfive})}
We now prove the left side of eqs. (\ref{eq:fivexfive}). 
The right-side one is shown by the same procedure. 
To prove the left-side equation, we first define 
an operator $A_\tau^{(t)}$ as follows
\begin{eqnarray}
\label{atau}
&&A_\tau^{(t)}=Q_\tau^{(t)\dagger} P_\tau^{(t)}Q_\tau^{(t)}. 
\end{eqnarray}
Then we need to show $A_\tau^{(t)}=P_0^{(t)}$. 
For this purpose, we now note that for $\tau =0$, 
\begin{eqnarray}
\label{azero}
&&A_0^{(t)}=P_0^{(t)}, 
\end{eqnarray} 
on account of $Q_0^{(t)}=1$. 
When we differentiate this operator with respect to $\tau$, 
we have 
\begin{eqnarray}
\label{adif}
\partial_\tau A_\tau^{(t)}&=&-Q_\tau^{(t)\dagger}\partial_\tau 
P_\tau^{(t)} P_\tau^{(t)}Q_\tau^{(t)}
+Q_\tau^{(t)\dagger} \partial_\tau P_\tau^{(t)}Q_\tau^{(t)}
\nonumber\\
&&-Q_\tau^{(t)\dagger} P_\tau^{(t)}\partial_\tau 
P_\tau^{(t)}Q_\tau^{(t)}\nonumber\\
&=&0,
\end{eqnarray} 
where we used the differential equation~(\ref{eq:fivexfour}) and
\begin{eqnarray}
&&P_\tau^{(t)}\partial_\tau P_\tau^{(t)}P_\tau^{(t)}=0.
\end{eqnarray}
From (\ref{azero}) and (\ref{adif}), we obtain 
$A_\tau^{(t)}=P_0^{(t)}$, i.e. 
\begin{eqnarray}
&&P_\tau^{(t)}=Q_\tau^{(t)} P_0^{(t)}Q_\tau^{(t)\dagger}. 
\end{eqnarray}

\subsubsection{eq. (\ref{eq:fivexthree})}
We here compute the Wilson line~(\ref{Wilsonline}) with a basis 
$\lc v,\bar{v}\rc$ along a closed curve, for which $U^1=U^0$,  
and show that it is equal to 
(\ref{eq:fivexthree}).  
Now the measure term is given by
\begin{eqnarray}
\label{measureint}
&&{\mathfrak L}_\xi=i\sum_{j}\left(v_j,\partial_\tau v_j\right)
+i\sum_k(\partial_\tau\overline v_k^\dagger,\overline v_k^\dagger).
\end{eqnarray}
We assume the basis $\lc v_j, \bar{v}_j\rc$ 
as $v_j|_{\tau=1}=v_j|_{\tau=0}$  
and $\bar{v}_k^\dagger |_{\tau=1}=\bar{v}_k^\dagger |_{\tau=0}$ and 
write the basis for $U^\tau$ as follows
\begin{eqnarray}
&&v_j=Q_\tau^{(t)}\sum_{l}v_l|_{\tau=0}(S^{-1})_{lj},\nonumber\\
&&\bar{v}_k^\dagger=\gamma_5 Q_\tau^{(2-t)} 
\gamma_5 \sum_{m}\bar{v}_m^\dagger |_{\tau=0}
(\bar{S}^{-1})_{mk}. 
\end{eqnarray}
where $S$ is the unitary-representation matrix of $Q_\tau^{(t)}$ 
on the space of left-handed fields and $\bar{S}$ is the one 
of $\gamma_5Q_\tau^{(2-t)}\gamma_5$ on the right-handed antifields and
$S|_{\tau=0}=1$ and $\bar{S}|_{\tau=0}=1$. 
Substituting this basis for (\ref{measureint}), we have 
\begin{eqnarray}
{\mathfrak L}_\xi&=&-i\Tr S^{-1}\partial_\tau S 
-i\Tr \bar{S}\partial_\tau (\bar{S}^{-1})\nonumber\\
&=&-i\left(\partial_\tau ln\det S +
\partial_\tau ln\det \bar{S}^{-1}\right), 
\end{eqnarray} 
where we used (\ref{eq:fivexfour}).  
Then the Wilson line along all closed curves is given by
\begin{eqnarray}
W&=&\det S|_{\tau=1}\det\nolimits^{-1}\bar{S}|_{\tau=1}
\nonumber\\
&=&\det(1-P_{-0}^{(t)}+P_{-0}^{(t)}Q_1^{(t)})
   \det\nolimits^{-1}(1-\overline P_{+0}^{(t)}
   +\overline P_{+0}^{(t)}\gamma_5Q_1^{(2-t)}\gamma_5)
\nonumber\\
&=&\det(1-P_{-0}^{(t)}+P_{-0}^{(t)}Q_1^{(t)})
   \det\nolimits^{-1}\lc\gamma_5(1-\overline P_{+0}^{(t)}
   +\overline P_{+0}^{(t)}\gamma_5Q_1^{(2-t)}\gamma_5)
\gamma_5\rc
\nonumber\\
&=&\det(1-P_{-0}^{(t)}+P_{-0}^{(t)}Q_1^{(t)})
   \det\nolimits^{-1}(1-P_{+0}^{(2-t)}
   +P_{+0}^{(2-t)}Q_1^{(2-t)}),
\end{eqnarray}
where we have written the signs of the chirality explicitly and   
$P_{+0}^{(2-t)}=P_+^{(2-t)}|_{U=U^0}$. 
From the third line to the fourth line in the above equation, 
we have utilized 
eq. (\ref{eq:twoxseven}). 
Noting the following equation:
\begin{eqnarray}
&&(1-P_{+0}^{(2-t)}+P_{+0}^{(2-t)}Q_1^{(2-t)})
(1-P_{-0}^{(2-t)}+P_{-0}^{(2-t)}Q_1^{(2-t)})=Q_1^{(2-t)},\nonumber\\
\end{eqnarray}
and $\det Q_\tau^{(t)}=1$, we have 
\begin{eqnarray}
\det\nolimits^{-1}(1-P_{+0}^{(2-t)}
   +P_{+0}^{(2-t)}Q_1^{(2-t)})=
\det(1-P_{-0}^{(2-t)}+P_{-0}^{(2-t)}Q_1^{(2-t)}),
\end{eqnarray}
and consequently obtain eq. (\ref{eq:fivexthree}).

\chapter{Conventions}\label{conventions}
This Appendix summarizes the $\gamma-$matrices
convention and the 
definitions of $C$, $P$ and $CP$ transformations.
We adopt the following convention of $\gamma$-matrices:
\begin{eqnarray}
   &&\{\gamma_\mu,\gamma_\nu\}=2\delta_{\mu\nu},\qquad
   \gamma_\mu^\dagger=\gamma_\mu,\qquad
   \gamma_5=-\gamma_1\gamma_2\gamma_3\gamma_4=\gamma_5^\dagger,
\nonumber\\
   &&\gamma_1^T=-\gamma_1,\qquad\gamma_2^T=\gamma_2,\qquad
   \gamma_3^T=-\gamma_3,\qquad\gamma_4^T=\gamma_4,\qquad
   \gamma_5^T=\gamma_5.\nonumber\\
\end{eqnarray}

The charge conjugation is defined by
\begin{eqnarray}
   &&\psi(x)\to-C^{-1}\overline\psi^T(x),\qquad
   \overline\psi(x)\to\psi^T(x)C,
\nonumber\\
   &&U(x,\mu)\to U^{\rm C}(x,\mu)=U(x,\mu)^*,
\end{eqnarray}
where the charge conjugation matrix~$C=\gamma_2\gamma_4$ satisfies
\begin{equation}
   C^\dagger C=1,\qquad C^T=-C,\qquad C\gamma_\mu C^{-1}=-\gamma_\mu^T,
   \qquad C\gamma_5C^{-1}=\gamma_5^T.
\end{equation}
Under this transformation, the kernel of Dirac operator transforms
as
\begin{equation}
   D(U^{\rm C})(x,y)=C^{-1}D(U)(x,y)^TC,
\end{equation}
where the transpose operation~$T$ acts not only on the matrices involved
but also on the arguments as $(x,y)\to(y,x)$.

The parity transformation is defined by
\begin{eqnarray}
   &&\psi(x)\to\gamma_4\psi(\bar x),\qquad
   \overline\psi(x)\to\overline\psi(\bar x)\gamma_4,
\nonumber\\
   &&U(x,\mu)\to U^{\rm P}(x,\mu)=\cases{
   U(\bar x-a\hat i,i)^{-1},&for $\mu=i$,\cr
   U(\bar x,4),& for $\mu=4$,\cr}
\end{eqnarray}
where
\begin{equation}
   \bar x=(-x_i,x_4),
\end{equation}
for $x=(x_i,x_4)$ ($i=1$, $2$, $3$). Under this,
\begin{equation}
   D(U^{\rm P})(x,y)=\gamma_4D(U)(\bar x,\bar y)\gamma_4.
\end{equation}

Finally, we define the CP transformation as
\begin{eqnarray}
   &&\psi(x)\to-W^{-1}\overline\psi^T(\bar x),\qquad
   \overline\psi(x)\to\psi^T(\bar x)W
\nonumber\\
   &&U_\mu(x)\to U^{\rm CP}_\mu(x)=\cases{
   {U_i(\bar x-a\hat i)^{-1}}^*,&for $\mu=i$,\cr
   U_4(\bar x)^*,& for $\mu=4$,\cr}
\end{eqnarray}
where
\begin{eqnarray}
   &&W=\gamma_2,\qquad W^\dagger W=1,
\nonumber\\
   &&W\gamma_\mu W^{-1}
   =\cases{\gamma_i^T,&for $\mu=i$,\cr
   -\gamma_4^T,& for $\mu=4$,\cr}\qquad
   W\gamma_5 W^{-1}=-\gamma_5^T,
\end{eqnarray}
and thus CP acts on the plaquette variables $U_{\mu\nu}(x)$ as
($\varphi_{ij}(\bar x-a\hat i-a\hat j)%
=U_j(\bar x-a\hat i-a\hat j)^*U_i(\bar x-a\hat i)^*$)
\begin{eqnarray}
   &&U_{ij}(x)
\nonumber\\
   &&\to U^{\rm CP}_{ij}(x)
   =\varphi_{ij}(\bar x-a\hat i-a\hat j)^{-1}
   U_{ij}(\bar x-a\hat i-a\hat j)^*
   \varphi_{ij}(\bar x-a\hat i-a\hat j),
\nonumber\\
   &&U_{i4}(x)\to U^{\rm CP}_{i4}(x)
   ={U_i(\bar x-a\hat i)^*}^{-1}
   {U_{i4}(\bar x-a\hat i)^{-1}}^*U_i(\bar x-a\hat i)^*.
\end{eqnarray}
Under this transformation, we have
\begin{equation}
   D(U^{\rm CP})(x,y)=W^{-1}D(U)(\bar x,\bar y)^TW.
\end{equation}
The CP transformation properties of various operators, which 
are used in this thesis, are given by 
\begin{eqnarray}
\label{variousCP}
&&H(U^{CP})=-W^{-1}(\gamma_5 H(U)\gamma_5)^T W,\qquad 
H(U^{CP})^2=W^{-1}H(U)^2W \nonumber\\
&&\Gamma_5(U^{CP})=-W^{-1}(\gamma_{5}\Gamma_{5}(U)
\gamma_{5})^T W,
\qquad\gamma_{5}\Gamma_{5}(U^{CP})=W^{-1}
(\gamma_{5}\Gamma_{5}(U))^{T}W,\nonumber\\
\end{eqnarray}  
and 
\begin{eqnarray}
\label{chiralCP}
   &&\gamma_5^{(t)}(U^{\rm CP})
   =-W^{-1}\overline\gamma_5^{(2-t)}(U)^TW,\qquad
   \overline\gamma_5^{(t)}(U^{\rm CP})=-W^{-1}\gamma_5^{(2-t)}(U)^TW,
\nonumber\\
   &&P_\pm^{(t)}(U^{\rm CP})=W^{-1}\overline P_\mp^{(2-t)}(U)^TW,
\qquad
   \overline P_\pm^{(t)}(U^{\rm CP})=W^{-1}P_\mp^{(2-t)}(U)^TW,
\nonumber\\
\end{eqnarray}
Throughout this thesis, the
transpose operation and the complex conjugation 
of an operator are
defined with respect to the corresponding kernel 
in position space.
Strictly speaking, the coordinates~$x$ 
in these expressions are 
replaced, for example, 
by $\bar x=(-x_i,x_4)$ under CP. We can forgo writing 
this explicitly,
because our final expressions always involve an 
integration 
over~$x$ and
$\sum_x=\sum_{\bar x}$.



\end{document}